\documentclass[10pt,journal,compsoc]{IEEEtran} % MP for now, this is the TMC template, but we can change it
\usepackage[numbers,sort&compress]{natbib}
\usepackage{amsmath,amssymb,amsfonts}
\usepackage{algorithmic}
\usepackage{graphicx}
\usepackage{textcomp}
\usepackage{xcolor}
\usepackage{caption}
\usepackage{subcaption}
\usepackage{dirtytalk}
\usepackage{multicol}
\usepackage{multirow}
\usepackage[normalem]{ulem}
\usepackage{lipsum} 
\usepackage{soul}
\usepackage{epsfig}
\usepackage{epstopdf}
\usepackage{graphics}
\usepackage{balance}
\usepackage{amssymb}
\usepackage{arydshln}
\usepackage{xspace}
\usepackage{enumitem}
\usepackage{graphicx}
\usepackage{colortbl}
\usepackage{hhline}
\usepackage{ragged2e}
\usepackage{array}
\usepackage{textgreek}

\usepackage[acronyms,nonumberlist,nopostdot,nomain,nogroupskip]{glossaries}
\newglossary[algh]{hidden}{acrh}{acnh}{Hidden Acronyms}

\newacronym{3gpp}{3GPP}{3rd Generation Partnership Project}
\newacronym{4g}{4G}{4th generation}
\newacronym{5g}{5G}{5th generation}
\newacronym{6g}{6G}{6th generation}
\newacronym{5gc}{5GC}{5G Core}
\newacronym{adc}{ADC}{Analog to Digital Converter}
\newacronym{aerpaw}{AERPAW}{Aerial Experimentation and Research Platform for Advanced Wireless}
\newacronym{ah}{AH}{Authentication Header}
\newacronym{ai}{AI}{Artificial Intelligence}
\newacronym{aimd}{AIMD}{Additive Increase Multiplicative Decrease}
\newacronym{am}{AM}{Acknowledged Mode}
\newacronym{amc}{AMC}{Adaptive Modulation and Coding}
\newacronym{amf}{AMF}{Access and Mobility Management Function}
\newacronym{aops}{AOPS}{Adaptive Order Prediction Scheduling}
\newacronym{api}{API}{Application Programming Interface}
\newacronym{apn}{APN}{Access Point Name}
\newacronym{aqm}{AQM}{Active Queue Management}
\newacronym{ausf}{AUSF}{Authentication Server Function}
\newacronym{avc}{AVC}{Advanced Video Coding}
\newacronym{awgn}{AGWN}{Additive White Gaussian Noise}
\newacronym{balia}{BALIA}{Balanced Link Adaptation Algorithm}
\newacronym{bbu}{BBU}{Base Band Unit}
\newacronym{bdp}{BDP}{Bandwidth-Delay Product}
\newacronym{ber}{BER}{Bit Error Rate}
\newacronym{bf}{BF}{Beamforming}
\newacronym{bler}{BLER}{Block Error Rate}
\newacronym{brr}{BRR}{Bayesian Ridge Regressor}
\newacronym{bsr}{BSR}{Buffer Status Report}
\newacronym{bs}{BS}{Base Station}
\newacronym{bpsk}{BPSK}{Binary Phase-shift keying}
\newacronym{bss}{BSS}{Business Support System}
\newacronym{ca}{CA}{Carrier Aggregation}
\newacronym{caas}{CaaS}{Connectivity-as-a-Service}
\newacronym{cb}{CB}{Code Block}
\newacronym{cc}{CC}{Congestion Control}
\newacronym{ccid}{CCID}{Congestion Control ID}
\newacronym{cco}{CC}{Carrier Component}
\newacronym{cdd}{CDD}{Cyclic Delay Diversity}
\newacronym{cdf}{CDF}{Cumulative Distribution Function}
\newacronym{cdn}{CDN}{Content Distribution Network}
\newacronym{cir}{CIR}{Channel Impulse Response}
\newacronym{cn}{CN}{Core Network}
\newacronym{codel}{CoDel}{Controlled Delay Management}
\newacronym{comac}{COMAC}{Converged Multi-Access and Core}
\newacronym{cord}{CORD}{Central Office Re-architected as a Datacenter}
\newacronym{cornet}{CORNET}{COgnitive Radio NETwork}
\newacronym{cosmos}{COSMOS}{Cloud Enhanced Open Software Defined Mobile Wireless Testbed for City-Scale Deployment}
\newacronym{cots}{COTS}{Commercial Off-the-Shelf}
\newacronym{cp}{C-plane}{Control Plane}
\newacronym{cpu}{CPU}{Central Processing Unit}
\newacronym{cqi}{CQI}{Channel Quality Information}
\newacronym{cr}{CR}{Cognitive Radio}
\newacronym{cran}{CRAN}{Cloud \gls{ran}}
\newacronym{crs}{CRS}{Cell Reference Signal}
\newacronym{csi}{CSI}{Channel State Information}
\newacronym{ct}{CT}{Cipher Text}
\newacronym{csirs}{CSI-RS}{Channel State Information - Reference Signal}
\newacronym{cu}{CU}{Central Unit}
\newacronym{d2tcp}{D$^2$TCP}{Deadline-aware Data center TCP}
\newacronym{d3}{D$^3$}{Deadline-Driven Delivery}
\newacronym{dac}{DAC}{Digital to Analog Converter}
\newacronym{dag}{DAG}{Directed Acyclic Graph}
\newacronym{darpa}{DARPA}{Defense Advanced Research Projects Agency}
\newacronym{das}{DAS}{Distributed Antenna System}
\newacronym{dash}{DASH}{Dynamic Adaptive Streaming over HTTP}
\newacronym{dc}{DC}{Dual Connectivity}
\newacronym{dccp}{DCCP}{Datagram Congestion Control Protocol}
\newacronym{dce}{DCE}{Direct Code Execution}
\newacronym{dci}{DCI}{Downlink Control Information}
\newacronym{dcl}{DCL}{Dear Colleague Letter}
\newacronym{dctcp}{DCTCP}{Data Center TCP}
\newacronym{dl}{DL}{Downlink}
\newacronym{dmr}{DMR}{Deadline Miss Ratio}
\newacronym{dmrs}{DMRS}{DeModulation Reference Signal}
\newacronym{drlcc}{DRL-CC}{Deep Reinforcement Learning Congestion Control}
\newacronym{drs}{DRS}{Discovery Reference Signal}
\newacronym{du}{DU}{Distributed Unit}
\newacronym{e2ap}{E2AP}{E2 Application Protocol}
\newacronym{e2e}{E2E}{end-to-end}
\newacronym{ecaas}{ECaaS}{Edge-Cloud-as-a-Service}
\newacronym{ecn}{ECN}{Explicit Congestion Notification}
\newacronym{edf}{EDF}{Earliest Deadline First}
\newacronym{em}{EM}{Electro-Magnetic}
\newacronym{embb}{eMBB}{Enhanced Mobile Broadband}
\newacronym{empower}{EMPOWER}{EMpowering transatlantic PlatfOrms for advanced WirEless Research}
\newacronym{enb}{eNB}{evolved Node Base}
\newacronym{endc}{EN-DC}{E-UTRAN-\gls{nr} \gls{dc}}
\newacronym{epc}{EPC}{Evolved Packet Core}
\newacronym{eps}{EPS}{Evolved Packet System}
\newacronym{es}{ES}{Edge Server}
\newacronym{esp}{ESP}{Encapsulating Security Payload}
\newacronym{etsi}{ETSI}{European Telecommunications Standards Institute}
\newacronym[firstplural=Estimated Times of Arrival (ETAs)]{eta}{ETA}{Estimated Time of Arrival}
\newacronym{eutran}{E-UTRAN}{Evolved Universal Terrestrial Access Network}
\newacronym{faas}{FaaS}{Function-as-a-Service}
\newacronym{fapi}{FAPI}{Functional Application Platform Interface}
\newacronym{fcc}{FCC}{Federal Communications Commission}
\newacronym{fdd}{FDD}{Frequency Division Duplexing}
\newacronym{fdm}{FDM}{Frequency Division Multiplexing}
\newacronym{fdma}{FDMA}{Frequency Division Multiple Access}
\newacronym{fed4fire}{FED4FIRE+}{Federation 4 Future Internet Research and Experimentation Plus}
\newacronym{fir}{FIR}{Finite Impulse Response}
\newacronym{fit}{FIT}{Future \acrlong{iot}}
\newacronym{fpga}{FPGA}{Field Programmable Gate Array}
\newacronym{fr2}{FR2}{Frequency Range 2}
\newacronym{fs}{FS}{Fast Switching}
\newacronym{fscc}{FSCC}{Flow Sharing Congestion Control}
\newacronym{ftp}{FTP}{File Transfer Protocol}
\newacronym{fw}{FW}{Flow Window}
\newacronym{ga128}{Ga}{Golay Sequence type A}
\newacronym{ge}{GE}{Gaussian Elimination}
\newacronym{glfsr}{GLFSR}{Galois Linear Feedback Shift Register}
\newacronym{gnb}{gNB}{Next Generation Node Base}
\newacronym{gold}{Gold}{Gold}
\newacronym{gop}{GOP}{Group of Pictures}
\newacronym{gpr}{GPR}{Gaussian Process Regressor}
\newacronym{gpu}{GPU}{Graphics Processing Unit}
\newacronym{gtp}{GTP}{GPRS Tunneling Protocol}
\newacronym{gtpc}{GTP-C}{GPRS Tunnelling Protocol Control Plane}
\newacronym{gtpu}{GTP-U}{GPRS Tunnelling Protocol User Plane}
\newacronym{gtpv2c}{GTPv2-C}{\gls{gtp} v2 - Control}
\newacronym{gw}{GW}{Gateway}
\newacronym{harq}{HARQ}{Hybrid Automatic Repeat reQuest}
\newacronym{hetnet}{HetNet}{Heterogeneous Network}
\newacronym{hh}{HH}{Hard Handover}
\newacronym{hol}{HOL}{Head-of-Line}
\newacronym{hqf}{HQF}{Highest-quality-first}
\newacronym{hss}{HSS}{Home Subscription Server}
\newacronym{http}{HTTP}{HyperText Transfer Protocol}
\newacronym{ia}{IA}{Initial Access}
\newacronym{iab}{IAB}{Integrated Access and Backhaul}
\newacronym{ic}{IC}{Incident Command}
\newacronym{ietf}{IETF}{Internet Engineering Task Force}
\newacronym{ifw}{IFW}{Interference Free Window}
\newacronym{imsi}{IMSI}{International Mobile Subscriber Identity}
\newacronym{imt}{IMT}{International Mobile Telecommunication}
\newacronym{iot}{IoT}{Internet of Things}
\newacronym{ip}{IP}{Internet Protocol}
\newacronym{ipsec}{IPsec}{Internet Protocol security}
\newacronym{iq}{IQ}{In-phase and Quadrature}
\newacronym{itu}{ITU}{International Telecommunication Union}
\newacronym{kpi}{KPI}{Key Performance Indicator}
\newacronym{kpm}{KPM}{Key Performance Metric}
\newacronym{kvm}{KVM}{Kernel-based Virtual Machine}
\newacronym{lan}{LAN}{Local Area Network}
\newacronym{los}{LOS}{Line-of-Sight}
\newacronym{ls}{LS}{Loosely Synchronised}
\newacronym{lsm}{LSM}{Link-to-System Mapping}
\newacronym{lstm}{LSTM}{Long Short Term Memory}
\newacronym{lte}{LTE}{Long Term Evolution}
\newacronym{lxc}{LXC}{Linux Container}
\newacronym{m2m}{M2M}{Machine to Machine}
\newacronym{mac}{MAC}{Medium Access Control}
\newacronym{macsec}{MACsec}{Media Access Control Security}
\newacronym{manet}{MANET}{Mobile Ad Hoc Network}
\newacronym{mano}{MANO}{Management and Orchestration}
\newacronym{mc}{MC}{Multi-Connectivity}
\newacronym{mcc}{MCC}{Mobile Cloud Computing}
\newacronym{mchem}{MCHEM}{Massive Channel Emulator}
\newacronym{mcs}{MCS}{Modulation and Coding Scheme}
\newacronym{mec}{MEC}{Multi-access Edge Computing}
\newacronym{mec2}{MEC}{Mobile Edge Cloud}
\newacronym{mfc}{MFC}{Mobile Fog Computing}
\newacronym{mi}{MI}{Mutual Information}
\newacronym{mib}{MIB}{Master Information Block}
\newacronym{miesm}{MIESM}{Mutual Information Based Effective SINR}
\newacronym{mimo}{MIMO}{Multiple Input, Multiple Output}
\newacronym{mgen}{MGEN}{Multi-Generator}
\newacronym{ml}{ML}{Machine Learning}
\newacronym{mlr}{MLR}{Maximum-local-rate}
\newacronym[plural=\gls{mme}s,firstplural=Mobility Management Entities (MMEs)]{mme}{MME}{Mobility Management Entity}
\newacronym{mmtc}{mMTC}{Massive Machine-Type Communications}
\newacronym{mmwave}{mmWave}{millimeter wave}
\newacronym{mpdccp}{MP-DCCP}{Multipath Datagram Congestion Control Protocol}
\newacronym{mptcp}{MPTCP}{Multipath TCP}
\newacronym{mr}{MR}{Maximum Rate}
\newacronym{mrdc}{MR-DC}{Multi \gls{rat} \gls{dc}}
\newacronym{mse}{MSE}{Mean Square Error}
\newacronym{mss}{MSS}{Maximum Segment Size}
\newacronym{mt}{MT}{Mobile Termination}
\newacronym{mtd}{MTD}{Machine-Type Device}
\newacronym{mtu}{MTU}{Maximum Transmission Unit}
\newacronym{mumimo}{MU-MIMO}{Multi-user \gls{mimo}}
\newacronym{mvno}{MVNO}{Mobile Virtual Network Operator}
\newacronym{nalu}{NALU}{Network Abstraction Layer Unit}
\newacronym{nas}{NAS}{Network Attached Storage}
\newacronym{nbiot}{NB-IoT}{Narrow Band IoT}
\newacronym{nfv}{NFV}{Network Function Virtualization}
\newacronym{nfvi}{NFVI}{Network Function Virtualization Infrastructure}
\newacronym{nic}{NIC}{Network Interface Card}
\newacronym{nlos}{NLOS}{Non-Line-of-Sight}
\newacronym{now}{NOW}{Non Overlapping Window}
\newacronym{nrdz}{NRDZ}{National Radio Dynamic Zone}
\newacronym{nsf}{NSF}{National Science Foundation}
\newacronym{nsm}{NSM}{Network Service Mesh}
\newacronym[type=hidden]{nr}{NR}{New Radio}
\newacronym{nrf}{NRF}{Network Repository Function}
\newacronym{nsa}{NSA}{Non Stand Alone}
\newacronym{nse}{NSE}{Network Slicing Engine}
\newacronym{nssf}{NSSF}{Network Slice Selection Function}
\newacronym{ntp}{NTP}{Network Time Protocol}
\newacronym{o2i}{O2I}{Outdoor to Indoor}
\newacronym{oai}{OAI}{OpenAirInterface}
\newacronym{oaicn}{OAI-CN}{\gls{oai} \acrlong{cn}}
\newacronym{oairan}{OAI-RAN}{\acrlong{oai} \acrlong{ran}}
\newacronym{oam}{OAM}{Operations, Administration and Maintenance}
\newacronym[plural=\gls{obu}s,firstplural=Onboard Units (OBUs)]{obu}{OBU}{Onboard Unit}
\newacronym{ofdm}{OFDM}{Orthogonal Frequency Division Multiplexing}
\newacronym{olia}{OLIA}{Opportunistic Linked Increase Algorithm}
\newacronym{omec}{OMEC}{Open Mobile Evolved Core}
\newacronym{onap}{ONAP}{Open Network Automation Platform}
\newacronym{onf}{ONF}{Open Networking Foundation}
\newacronym{onos}{ONOS}{Open Networking Operating System}
\newacronym{oom}{OOM}{\gls{onap} Operations Manager}
\newacronym{opnfv}{OPNFV}{Open Platform for \gls{nfv}}
\newacronym[type=hidden]{o-ran}{O-RAN}{Open \gls{ran}}
\newacronym{orbit}{ORBIT}{Open-Access Research Testbed for Next-Generation Wireless Networks}
\newacronym{os}{OS}{Operating System}
\newacronym{osc}{OSC}{O-RAN Software Community}
\newacronym{osm}{OSM}{Open Street Map}
\newacronym{oss}{OSS}{Operations Support System}
\newacronym{pa}{PA}{Position-aware}
\newacronym{pase}{PASE}{Prioritization, Arbitration, and Self-adjusting Endpoints}
\newacronym{pawr}{PAWR}{Platforms for Advanced Wireless Research}
\newacronym{pbch}{PBCH}{Physical Broadcast Channel}
\newacronym{pcef}{PCEF}{Policy and Charging Enforcement Function}
\newacronym{pcfich}{PCFICH}{Physical Control Format Indicator Channel}
\newacronym{pcrf}{PCRF}{Policy and Charging Rules Function}
\newacronym{pdcch}{PDCCH}{Physical Downlink Control Channel}
\newacronym{pdcp}{PDCP}{Packet Data Convergence Protocol}
\newacronym{pdsch}{PDSCH}{Physical Downlink Shared Channel}
\newacronym{pdu}{PDU}{Packet Data Unit}
\newacronym{pdp}{PDP}{Power Delay Profile}
\newacronym{pf}{PF}{Proportional Fair}
\newacronym{pgw}{PGW}{Packet Gateway}
\newacronym{phich}{PHICH}{Physical Hybrid ARQ Indicator Channel}
\newacronym{phy}{PHY}{Physical}
\newacronym{pl}{PL}{Path Loss}
\newacronym{pt}{PT}{Plain Text}
\newacronym{ptp}{PTP}{Precision Time Protocol}
\newacronym{pmch}{PMCH}{Physical Multicast Channel}
\newacronym{pmi}{PMI}{Precoding Matrix Indicators}
\newacronym{powder}{POWDER}{Platform for Open Wireless Data-driven Experimental Research}
\newacronym{ppo}{PPO}{Proximal Policy Optimization}
\newacronym{ppp}{PPP}{Poisson Point Process}
\newacronym{prach}{PRACH}{Physical Random Access Channel}
\newacronym{prb}{PRB}{Physical Resource Block}
\newacronym{psnr}{PSNR}{Peak Signal to Noise Ratio}
\newacronym{pss}{PSS}{Primary Synchronization Signal}
\newacronym{pucch}{PUCCH}{Physical Uplink Control Channel}
\newacronym{pusch}{PUSCH}{Physical Uplink Shared Channel}
\newacronym{qam}{QAM}{Quadrature Amplitude Modulation}
\newacronym{qci}{QCI}{\gls{qos} Class Identifier}
\newacronym{qoe}{QoE}{Quality of Experience}
\newacronym{qos}{QoS}{Quality of Service}
\newacronym{qtgui}{QT-GUI}{QT Graphical User Interface}
\newacronym{quic}{QUIC}{Quick UDP Internet Connections}
\newacronym{rach}{RACH}{Random Access Channel}
\newacronym{ran}{RAN}{Radio Access Network}
\newacronym[firstplural=Radio Access Technologies (RATs)]{rat}{RAT}{Radio Access Technology}
\newacronym{rcn}{RCN}{Research Coordination Network}
\newacronym{rec}{REC}{Radio Edge Cloud}
\newacronym{red}{RED}{Random Early Detection}
\newacronym{renew}{RENEW}{Reconfigurable Eco-system for Next-generation End-to-end Wireless}
\newacronym{rf}{RF}{Radio Frequency}
\newacronym{rfc}{RFC}{Request for Comments}
\newacronym{rfr}{RFR}{Random Forest Regressor}
\newacronym{ric}{RIC}{\gls{ran} Intelligent Controller}
\newacronym{rlc}{RLC}{Radio Link Control}
\newacronym{rlf}{RLF}{Radio Link Failure}
\newacronym{rlnc}{RLNC}{Random Linear Network Coding}
\newacronym{rmse}{RMSE}{Root Mean Squared Error}
\newacronym{rnis}{RNIS}{Radio Network Information Service}
\newacronym{rr}{RR}{Round Robin}
\newacronym{rrc}{RRC}{Radio Resource Control}
\newacronym{rrm}{RRM}{Radio Resource Management}
\newacronym{rru}{RRU}{Remote Radio Unit}
\newacronym{rs}{RS}{Remote Server}
\newacronym{rsrp}{RSRP}{Reference Signal Received Power}
\newacronym{rsrq}{RSRQ}{Reference Signal Received Quality}
\newacronym{rss}{RSS}{Received Signal Strength}
\newacronym{rssi}{RSSI}{Received Signal Strength Indicator}
\newacronym{rsu}{RSU}{Road-Side Unit}
\newacronym{rtt}{RTT}{Round Trip Time}
\newacronym{ru}{RU}{Radio Unit}
\newacronym{rw}{RW}{Receive Window}
\newacronym{rx}{RX}{Receiver}
\newacronym{s1ap}{S1AP}{S1 Application Protocol}
\newacronym{sa}{SA}{Security Association}
\newacronym{sack}{SACK}{Selective Acknowledgment}
\newacronym{sap}{SAP}{Service Access Point}
\newacronym{sc2}{SC2}{Spectrum Collaboration Challenge}
\newacronym{scef}{SCEF}{Service Capability Exposure Function}
\newacronym{sch}{SCH}{Secondary Cell Handover}
\newacronym{scoot}{SCOOT}{Split Cycle Offset Optimization Technique}
\newacronym{sctp}{SCTP}{Stream Control Transmission Protocol}
\newacronym{sdap}{SDAP}{Service Data Adaptation Protocol}
\newacronym{sd}{SD}{Standard Deviation}
\newacronym{sdk}{SDK}{Software Development Kit}
\newacronym{sdm}{SDM}{Space Division Multiplexing}
\newacronym{sdma}{SDMA}{Spatial Division Multiple Access}
\newacronym{sdn}{SDN}{Software-defined Networking}
\newacronym{sdr}{SDR}{Software-defined Radio}
\newacronym{seba}{SEBA}{SDN-Enabled Broadband Access}
\newacronym{sgsn}{SGSN}{Serving GPRS Support Node}
\newacronym{sgw}{SGW}{Service Gateway}
\newacronym{si}{SI}{Study Item}
\newacronym{sib}{SIB}{Secondary Information Block}
\newacronym{sinr}{SINR}{Signal to Interference plus Noise Ratio}
\newacronym{sip}{SIP}{Session Initiation Protocol}
\newacronym{siso}{SISO}{Single Input, Single Output}
\newacronym{sla}{SLA}{Service Level Agreement}
\newacronym{sm}{SM}{Saturation Mode}
\newacronym{smf}{SMF}{Session Management Function}
\newacronym{smo}{SMO}{Service Management and Orchestration}
\newacronym{sms}{SMS}{Short Message Service}
\newacronym{smsgmsc}{SMS-GMSC}{\gls{sms}-Gateway}
\newacronym{snr}{SNR}{Signal-to-Noise-Ratio}
\newacronym{son}{SON}{Self-Organizing Network}
\newacronym{sptcp}{SPTCP}{Single Path TCP}
\newacronym{srb}{SRB}{Service Radio Bearer}
\newacronym{srn}{SRN}{Standard Radio Node}
\newacronym{srs}{SRS}{Sounding Reference Signal}
\newacronym{ss}{SS}{Synchronization Signal}
\newacronym{sss}{SSS}{Secondary Synchronization Signal}
\newacronym{st}{ST}{Spanning Tree}
\newacronym{svc}{SVC}{Scalable Video Coding}
\newacronym{tb}{TB}{Transport Block}
\newacronym{tcp}{TCP}{Transmission Control Protocol}
\newacronym{tdd}{TDD}{Time Division Duplexing}
\newacronym{tdm}{TDM}{Time Division Multiplexing}
\newacronym{tdma}{TDMA}{Time Division Multiple Access}
\newacronym{tfl}{TfL}{Transport for London}
\newacronym{tfrc}{TFRC}{TCP-Friendly Rate Control}
\newacronym{tft}{TFT}{Traffic Flow Template}
\newacronym{tgen}{TGEN}{Traffic Generator}
\newacronym{tip}{TIP}{Telecom Infra Project}
\newacronym{tls}{TLS}{Transport Layer Security}
\newacronym{tm}{TM}{Transparent Mode}
\newacronym{to}{TO}{Telco Operator}
\newacronym{toa}{ToA}{Time of Arrival}
\newacronym{tr}{TR}{Technical Report}
\newacronym{trp}{TRP}{Transmitter Receiver Pair}
\newacronym{ts}{TS}{Technical Specification}
\newacronym{tti}{TTI}{Transmission Time Interval}
\newacronym{ttt}{TTT}{Time-to-Trigger}
\newacronym{tx}{TX}{Transmitter}
\newacronym{uas}{UAS}{Unmanned Aerial System}
\newacronym{uav}{UAV}{Unmanned Aerial Vehicle}
\newacronym{udm}{UDM}{Unified Data Management}
\newacronym{udp}{UDP}{User Datagram Protocol}
\newacronym{udr}{UDR}{Unified Data Repository}
\newacronym{ue}{UE}{User Equipment}
\newacronym{uhd}{UHD}{\gls{usrp} Hardware Driver}
\newacronym{ul}{UL}{Uplink}
\newacronym{um}{UM}{Unacknowledged Mode}
\newacronym{uml}{UML}{Unified Modeling Language}
\newacronym{up}{U-plane}{User Plane}
\newacronym{upa}{UPA}{Uniform Planar Array}
\newacronym{upf}{UPF}{User Plane Function}
\newacronym{urllc}{URLLC}{Ultra Reliable and Low Latency Communication}
\newacronym{usa}{U.S.}{United States}
\newacronym{usim}{USIM}{Universal Subscriber Identity Module}
\newacronym{usrp}{USRP}{Universal Software Radio Peripheral}
\newacronym{utc}{UTC}{Urban Traffic Control}
\newacronym{vim}{VIM}{Virtualization Infrastructure Manager}
\newacronym{vm}{VM}{Virtual Machine}
\newacronym{vnf}{VNF}{Virtual Network Function}
\newacronym{volte}{VoLTE}{Voice over \gls{lte}}
\newacronym{voltha}{VOLTHA}{Virtual OLT HArdware Abstraction}
\newacronym{vr}{VR}{Virtual Reality}
\newacronym{vran}{vRAN}{Virtualized \gls{ran}}
\newacronym{vss}{VSS}{Video Streaming Server}
\newacronym{wbf}{WBF}{Wired Bias Function}
\newacronym{wf}{WF}{Wired-first}
\newacronym{wi}{WI}{Wireless InSite}
\newacronym{wlan}{WLAN}{Wireless Local Area Network}
\newacronym{pnf}{PNF}{Physical Network Function}
\newacronym{drl}{DRL}{Deep Reinforcement Learning}
\newacronym{mtc}{MTC}{Machine-type Communications}
\newacronym{v2x}{V2X}{Vehicle-to-everything}
\newacronym{cast}{\textit{CaST}}{Channel emulation generator and Sounder Toolchain}
\newacronym{arc}{ARC}{Aerial Research Cloud}
\newacronym{dsp}{DSP}{Digital Signal Processing}
\newacronym{ota}{OTA}{Over-the-Air}
\newacronym{bom}{BoM}{Bill of Materials}
\newacronym{frand}{FRAND}{fair, reasonable, and non-discriminatory}
\newacronym{ipc}{IPC}{Inter-Process Communications}
\newacronym{uci}{UCI}{Uplink Control Indication}
\newacronym{rdma}{RDMA}{remote direct memory access}
\newacronym{oran}{Open RAN}{Open Radio Access Network}

\newcommand{\rev}[1]{{\color{red}#1}}

\def\BibTeX{{\rm B\kern-.05em{\sc i\kern-.025em b}\kern-.08em
    T\kern-.1667em\lower.7ex\hbox{E}\kern-.125emX}}
    
\begin{document}

\title{Securing O-RAN Open Interfaces\\
}

\author{
\IEEEauthorblockN{
Joshua Groen, 
Salvatore D'Oro, 
Utku Demir, 
Leonardo Bonati, 
Davide Villa, \\
Michele Polese,
Tommaso Melodia,
Kaushik Chowdhury
}
\thanks{The authors are with the Institute for the Wireless Internet of Things, Northeastern University, Boston, MA, USA. E-mail: \{groen.j, s.doro, u.demir, l.bonati, villa.d, m.polese, t.melodia, k.chowdhury\}@northeastern.edu.}
\thanks{This article is based upon work partially supported by Qualcomm, Inc. and by the U.S.\ National Science Foundation under grants CNS-1925601, CNS-2112471, CNS-2117814, and CNS-2120447.}
%\thanks{\hl{Michele: I would also ack Open6G, thoughts?}}
% \textit{Institute for the Wireless Internet of Things}, Northeastern University, Boston, MA, USA \\
% \IEEEauthorblockN{
% \IEEEauthorrefmark{1}\{last.f\}@northeastern.edu,
% \{f.last\}@northeastern.edu} \\ 
}

\maketitle

\begin{abstract}

%A promising vision for the emerging next generation of cellular networks  is one that embraces openness, intelligence, virtualization, and distributed computing.  The Open Radio Access Network framework is making significant strides toward these goals and is already seeing prototype deployments in academia and industry. While there is general consensus that this technology may disrupt the status quo by eliminating vendor lock-ins, there are serious questions about the security implications in such disaggregated networks. Indeed, securing data and controlling interfaces must be a core consideration in the design of O-RAN and cost/benefit tradeoffs need to be rigorously analyzed. In this paper, we undertake the first systematic study on the impact of encryption on two critical O-RAN interfaces. The E2 interface, connecting the base station to a near-real time radio intelligence controller, and the Open Fronthaul connecting the O-RU to the O-DU. We conduct this study using a full stack O-RAN compliant implementation in the Colosseum network emulator as well as in a private OTA 5G network. The contributions of this paper include quantitative measurements of added latency and reduced throughput due to encryption protocols. We also present four key principles to building security by design in Open RAN systems.

The next generation of cellular networks will be characterized by openness, intelligence, virtualization, and distributed computing. The \gls{oran} framework represents a significant leap toward realizing these ideals, with prototype deployments taking place in both academic and industrial domains. While it holds the potential to disrupt the established vendor lock-ins, Open RAN's disaggregated nature raises critical security concerns. Safeguarding data and securing interfaces must be integral to Open RAN's design, demanding meticulous analysis of cost/benefit tradeoffs.

In this paper, we embark on the first comprehensive investigation into the impact of encryption on two pivotal \gls{oran} interfaces: the E2 interface, connecting the base station with a near-real-time RAN Intelligent Controller, and the Open Fronthaul, connecting the Radio Unit to the Distributed Unit. Our study leverages a full-stack O-RAN ALLIANCE compliant implementation within the Colosseum network emulator and a production-ready Open RAN and 5G-compliant private cellular network. This research contributes quantitative insights into the latency introduced and throughput reduction stemming from using various encryption protocols. Furthermore, we present four fundamental principles for constructing security by design within Open RAN systems, offering a roadmap for navigating the intricate landscape of Open RAN security.

\end{abstract}


%%% uncomment the following to make the arXiv acceptance notice appear
\begin{picture}(0,0)(10,-450)
\put(0,0){
\put(0,0){\footnotesize This paper has been accepted for publication on IEEE Transactions on Mobile Computing, DOI 10.1109 / TMC.2024.3393430.}
\put(0,-10){
\scriptsize \textcopyright~2024 IEEE. Personal use of this material is permitted. Permission from IEEE must be obtained for all other uses, in any current or future media, including}
\put(0, -17){
\scriptsize reprinting/republishing this material for advertising or promotional purposes, creating new collective works, for resale or redistribution to servers or lists,}
\put(0, -24){
\scriptsize or reuse of any copyrighted component of this work in other works.}
}
\end{picture}

\section{Introduction}

\begin{figure}[tb]
    \centering
    \includegraphics[width=\linewidth]{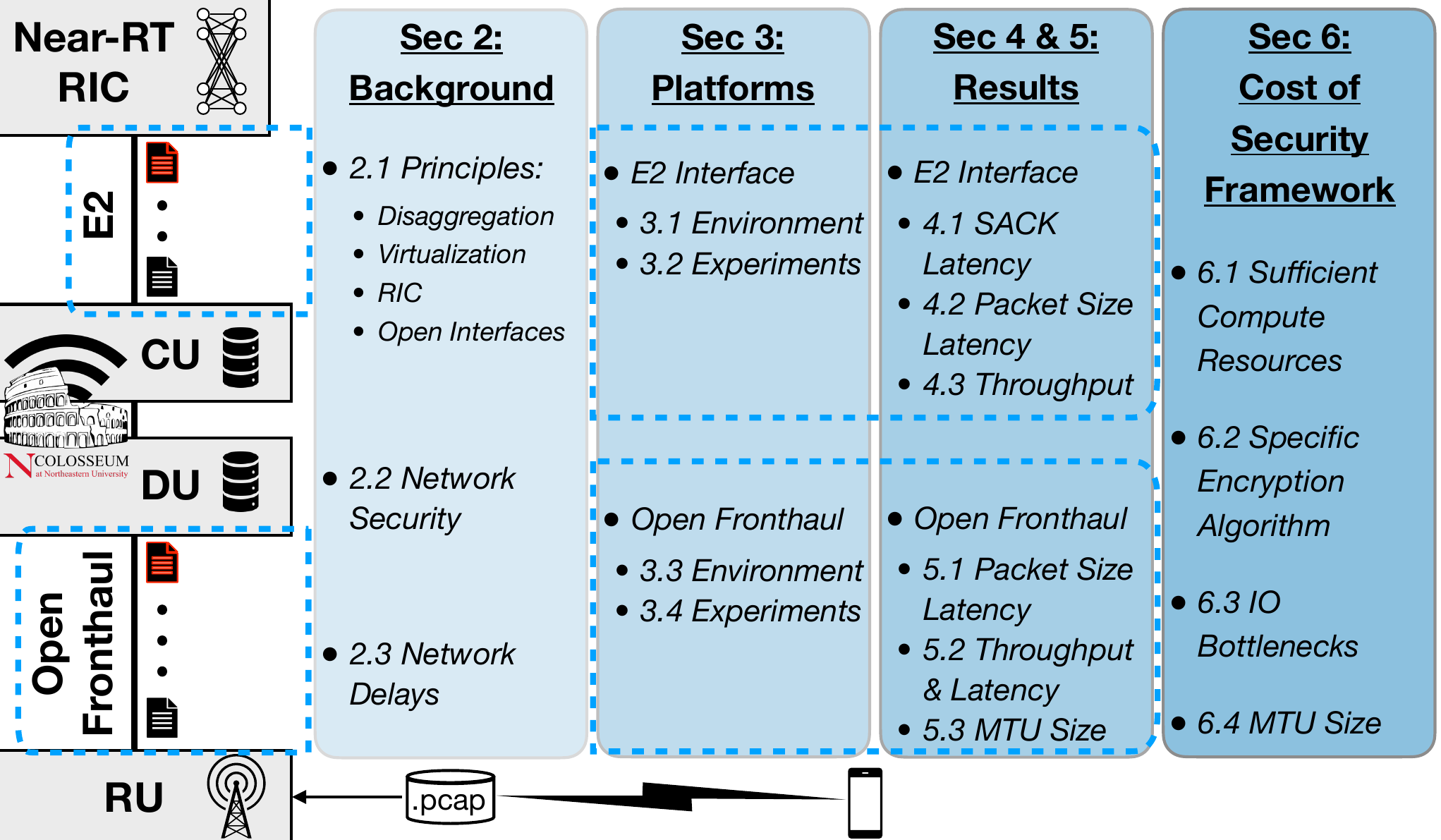}
    %\caption{\rev{Replace this figure.} E2 interface connects near-RT RIC to eNB and nearly every component in the next generation distributed gNB.}
    \caption{Overview of the study, focusing on securing the O-RAN architecture's open interfaces. %\rev{figure labels needs a change + Is the base station icon by the RU correct? + Fonts?}
    }
    \label{fig:E2interface}
\end{figure}

The Open Radio Access Network (Open RAN) paradigm and its embodiment in the O-RAN ALLIANCE specifications~\cite{polese2023understanding_official} aim to transform the 5G (and beyond) ecosystem via open, intelligent, virtualized, and fully inter-operable RANs. The O-RAN architecture separates itself from legacy blackbox and monolithic RAN architectures by adopting a more flexible approach where base stations, e.g., \glspl{gnb}, are converted into disaggregated and virtualized components that are connected through open and standardized interfaces \cite{OranWG1}. For example, Fig.~\ref{fig:E2interface} shows the Open Fronthaul connecting the RU and DU and the E2 interface connecting the DU and CU to the near-RT RIC. This paradigm shift is a potential enabler of data-driven optimization, closed-loop control, and automation \cite{thaliath2022predictive}, thus making it possible to break the stagnant vendor lock-in of closed networking architectures used in 4G legacy systems. 
The introduction of open interfaces to interconnect disaggregated RAN components paves the way to intelligent, flexible, and scalable cellular architectures. However, this effectively opens up the network and allow access to its elements via software, extending the threat surface and exposing the network to a variety of vulnerabilities and attacks~\cite{shen2022security, ramezanpour2022intelligent}. 

%\rev{MP: do we really want to say this?} Alongside traditional telecommunication operators who are active members of the O-RAN ALLIANCE, the US Department of Defense is actively promoting open interfaces in both the RAN and the 5G Core that will pave the way toward sustainable cellular networks, allow for more competition and boost innovation~\cite{DoD}. 
%KRC- ORAN or O-RAN? Need to be consistent.

In fact, one of the major threats to O-RAN open interfaces arises from improper or missing ciphering of the data sent across them \cite{abdalla2022toward, ramezanpour2022intelligent}. O-RAN interfaces transport potentially sensitive user data and network telemetry and control, which need protection against data tampering and eavesdropping. Similarly, the introduction of \glspl{ric}—software entities hosting \gls{ai} algorithms executed through applications known as xApps and rApps—renders the network susceptible to various adversarial attacks. These attacks aim to manipulate the \gls{ai} towards inefficient control policies or decisions that could potentially degrade network performance~\cite{groen2023implementing} and allocate RAN resources unfairly~\cite{10.1145/3465481.3470080}. Additionally, adversaries might attempt to bypass authentication procedures, thereby authorizing the execution of malicious software applications~\cite{10056720, 10.1145/3495243.3558259}.

Security is a crucial aspect of any system as it influences its adoption and utilization. The success of O-RAN will inevitably rely upon its security framework, procedures, and defense mechanisms.
For this reason, the O-RAN ALLIANCE, industry, government organizations, and academia alike have put significant effort in laying out best practices and identifying threats and their countermeasures~\cite{ericcson, OranWG3, DoD, shen2022security, abdalla2022toward, ramezanpour2022intelligent}. For example, \emph{O-RAN WG11: Security Work Group} has developed an extensive threat analysis for O-RAN systems and recommendations to secure them. More specifically, \emph{O-RAN WG3: The Near-Real-Time RIC and E2 Interface Work Group} is working on identifying threats and guidance to secure the E2 open interface with confidentiality, integrity, replay protection, and data origin authentication mechanisms~\cite{OranWG3}. %While the O-RAN ALLIANCE does not require any security mechanism for the Open Fronthaul, Cho and Sergeev propose using MACsec to prevent a host of threats to this interface, including: inserting traffic, network intrusion, and man-in-the middle attacks to monitor, modify, or inject control messages~\cite{10.1145/3465481.3470080}. 
In contrast, at the time of writing, the widely adopted Open Fronthaul interface O-RAN standards call for no encryption mechanism because of the high bandwidth and strict latency requirements. Nonetheless,
% Although the O-RAN ALLIANCE currently specifies that no encryption should be implemented over the Open Fronthaul because of the high bandwidth and strict latency requirements, 
there are works in the literature that suggest using \gls{macsec} for this interface~\cite{10.1145/3465481.3470080, 9604996}. \gls{macsec} can be used to prevent a host of threats to this interface, including: inserting traffic, network intrusion, and man-in-the-middle attacks. %to monitor, modify, or inject control messages. 
However, prior work does not consider an actual implementation of \gls{macsec} for the Open Fronthaul nor analyze its overhead cost. %\cite{ramezanpour2022intelligent} takse a different approach, proposing the adoption of a zero trust architecture (ZTA) for Open RAN. ZTA assumes network devices are untrusted and takes a data-centric approach that continuously assesses trust and risk granting the least privilege necessary for any specific user and task. 

\noindent
\textbf{Motivation.} 
Although the examples above demonstrate traction and desire to make O-RAN secure by addressing a variety of threats, to the best of our knowledge there has been no systematic study to identify and measure the cost that security has on O-RAN open interfaces. It is vital that an informed and risk-based approach is taken to adequately address security concerns in O-RAN, while recognizing that any method for enhancing security, such as adding encryption, comes at a performance cost \cite{ericcson}. \textcolor{black}{Our goal in this paper is to (i) understand if such a cost is tolerable (motivating a rapid adoption of security protocols that can be used without impacting performance and normal operations of the network); and (ii) identify which elements contribute the most to such costs informing system architects on how to design security systems for O-RAN that are practical and sustainable.}

To derive practical insights, it is essential for security studies to rely on empirical data obtained from the execution of security algorithms on O-RAN hardware and software components. This approach enables accurate measurement of how security mechanisms impact resource utilization, processing latency, and data rates. For this reason, we leverage the broad and public availability of O-RAN testbeds~\cite{bonati2021colosseum, bonati2022openran, bonati2021scope, polese2022colo, villa2023x5g} to thoroughly test and analyze the effects of encryption on a variety of O-RAN interfaces. 
We distinguish between two primary categories of interfaces within our study, offering specific examples of each to provide a clear framework for our analysis. This approach allows us to delve into the unique characteristics and requirements of each interface type, aiding in a more precise evaluation of the impact of security measures.

\begin{enumerate}[wide, labelindent=0pt]
    \item \textbf{\gls{ric} Data Interfaces:} This category comprises interfaces that the \glspl{ric} utilize for both receiving and transmitting data. The O-RAN architecture features two \glspl{ric}, executing control loops with a near-real-time scale (10 ms to 1 s) and a non-real-time scale (beyond 1 s). Notable examples of interfaces for the \glspl{ric} include E2, O1, and A1. For our analysis, we specifically focus on the impact of adding \gls{ipsec} to the E2 interface by extending our prior works and findings in \cite{groenCost, groen2023implementing}, as it plays a pivotal role in facilitating the exchange of packets between the near-RT \gls{ric} and the \glspl{cu}/\glspl{du}. %This interface is designed to handle telemetry and control transfer within a time frame ranging from 10 ms to 1 s.  
    \item \textbf{Interfaces Enabling \gls{gnb} Disaggregation:} The second category encompasses interfaces that are essential for supporting the disaggregation of \glspl{gnb}. One of the key interfaces within this category is the Open Fronthaul interface for the 7.2 split of the 3GPP stack. This interface---defined by the O-RAN ALLIANCE---serves as a critical link between the \gls{ru} and the \gls{du}, managing the transmission of time-sensitive data in substantial volumes, such as \gls{iq} samples. For this class, we examine the cost of securing the Open Fronthaul interface with \gls{macsec}.
\end{enumerate}

The main contributions of our work are
% can be summarized 
as follows:
%%%% OLD
% \begin{itemize}
%     \item We perform the first ever experimental analysis of the effects of adding O-RAN-compliant encryption to the O-RAN E2 and Open Fronthaul interfaces using Colosseum and the X-Mili testbeds. 
%     [Section \ref{par:system_overview}]
%    \item We evaluate the impact that security has on latency and validate at-scale a theoretical framework for calculating the total network delay when adding security protocols to O-RAN-based distributed functional units. [Section \ref{types_of_delay}]. We extend these results by developing a general framework for understanding the cost of encryption in O-RAN which we hope will enable researchers and engineers to build secure-by-design O-RAN systems. [Section \ref{framework}]
% \rev{
%     \item We develop two separate O-RAN emulation environments and publicly release a set of tools \rev{[are we also releasing any dataset?]} used to analyze the impact of adding security to Open Interfaces. [Section \ref{setup}]
%     \item We report for the first time an experimental analysis of the effects of adding security with \gls{macsec} to the O-RAN Open Fronthaul interface using two separate O-RAN emulation environments. [Section \ref{results fronthaul}]
%     \item We validate and extend previous studies analyzing the impact of adding security to the E2 interface with \gls{ipsec}. [Section \ref{results E2}]
%     \item We extend these results by developing four key principles that system designers should use to build future O-RAN systems with security by design. [Section \ref{framework}]
%     }
% \end{itemize}

%%% NEW
\begin{itemize}[wide, labelindent=0pt]
    \item We conduct the first-ever experimental analysis of adding O-RAN-compliant encryption to the O-RAN E2 and Open Fronthaul interfaces using Colosseum and a private 5G O-RAN-compliant testbed. Specifically, we extend our previous study~\cite{groenCost} that analyzes the impact of adding security to the E2 interface with \gls{ipsec} and report, for the first time, an experimental analysis of the effects of adding security with \gls{macsec} to the O-RAN Open Fronthaul interface.
   %ORIGINAL
   %\item We evaluate the impact that security has on latency and validate at-scale a theoretical framework for calculating the total network delay when adding security protocols to %O-RAN-based 
   %distributed functional units. We extend these results by developing a general framework for understanding the cost of encryption in O-RAN with the goal of enabling researchers and engineers to build secure-by-design O-RAN systems.
   %which we hope will enable researchers and engineers to build secure-by-design O-RAN systems.
   \item \textcolor{black}{We showcase the validity of our results through live experimentation that sheds light into the performance of O-RAN interfaces when the proposed security measures are deployed. Additionally, we validate at-scale a theoretical framework for calculating the total network delay when adding security protocols to distributed functional units}. We extend these results by developing a general framework for understanding the cost of encryption in O-RAN with the goal of enabling researchers and engineers to build secure-by-design O-RAN systems.
   %which we hope will enable researchers and engineers to build secure-by-design O-RAN systems.
   \item We develop two separate O-RAN emulation environments and \textcolor{black}{will publicly release the set of tools and datasets used to analyze the impact of adding security to Open Interfaces upon acceptance of this paper}. 
   \item We derive insights and identify four key principles that system designers should be aware of to build future O-RAN systems that are secure by design. These principles are: sufficient compute resources must be provisioned, specific protocol implementations and encryption algorithms matter greatly, user space and kernel space I/O bottlenecks must be addressed, and the network \gls{mtu} size should be optimized. 
\end{itemize}

% \rev{Of particular importance is the O-RAN Open Fronthaul interface. This vital interface connects the disaggregated RU and DU. Currently, the O-RAN ALLIANCE specifies that no security should be implemented over the Open Fronthaul because of the high bandwidth and strict latency requirements. Other works have suggested using \gls{macsec} to secure the Open Fronthaul though they did not impliment \gls{macsec} or analyze its cost \cite{10.1145/3465481.3470080, 9604996}. In this paper, we present the results of a thorough analsys of the impact of adding \gls{macsec} to the Open Fronthaul interface. We also examine the cost of securing other interfaces such as the E2 interface with \gls{ipsec} as demonstrated in \cite{groenCost}. Based on our systematic study of \gls{macsec} and \gls{ipsec} implemented in real systems emulating an O-RAN environment, we present four key principles that must be considered when adding security to any of the Open Interfaces.}

The rest of the paper's organization is shown in Fig.~\ref{fig:E2interface}. Section \ref{s: background} gives a brief overview of key O-RAN principles. Section \ref{s: environment} describes our emulation environments.  Our experimental procedures and results are detailed in Section \ref{s:E2} for the E2 interface and in Section \ref{s:OF} for the Open Fronthaul interface. We provide additional analysis of our results in Section \ref{s:analysis} and conclude the paper in Section \ref{s:conclusion}.

\section{Background}\label{s: background}
\begin{figure}
    \centering
    \includegraphics[width=.9\linewidth]{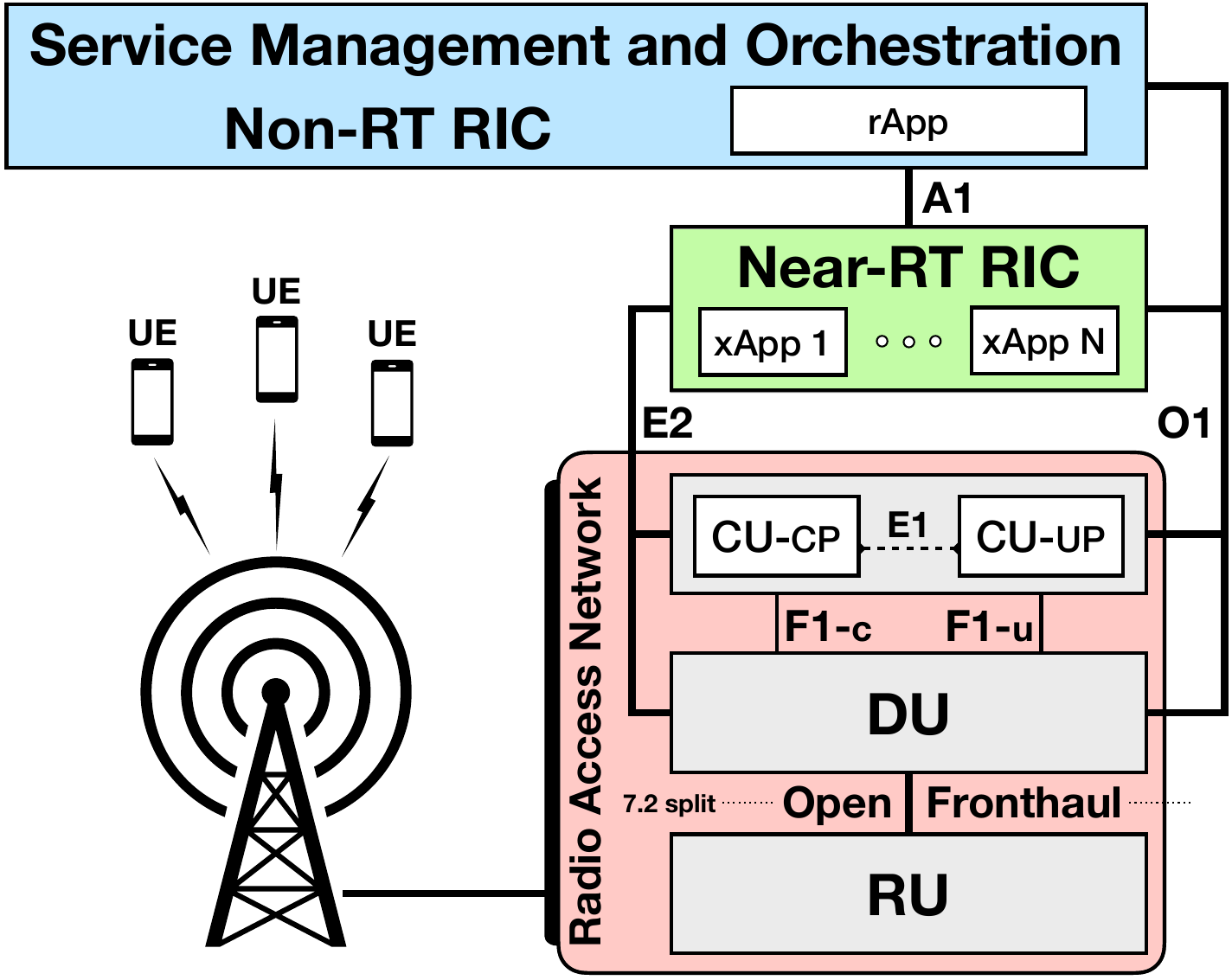}
    \caption{Overall O-RAN architecture showing the 7.2x split with disaggregated \gls{du}, \gls{ru}, and \gls{cu}. The Near-RT and Non-RT \glspl{ric} house \gls{ml} models to provide closed-loop RAN control under different time scales.}
    \label{fig:ORAN_arch}
\end{figure}

The goal of this section is to provide %basic notions and 
background that will be useful later, when we delve into the details of the security assessment study. We first provide a brief overview of O-RAN shown in Fig.~\ref{fig:ORAN_arch}, and its foundational principles (Sec.~\ref{sec:back:oran}), and then introduce network security protocols that will be used in this work (Sec.~\ref{sec:back:security}). Finally, we give an overview of the types of delay in packet switched networks (Sec.~\ref{types_of_delay}).

%\rev{Do we want to provide a figure that shows the complete picture of O-RAN? In in the current version of Fig.1, we zoom into the focused area. MP: yes, good idea!}.  
% It is important to understand these principles and the basic architecture of the O-RAN system to understand the key role of the E2 interface.

\subsection{O-RAN Principles and Architecture} \label{sec:back:oran}

\noindent $\bullet$~\textbf{Disaggregation and Virtualization:} O-RAN embraces disaggregation principles by effectively splitting base stations into multiple functional units, namely the \gls{cu}, the \gls{du}, and the \gls{ru}. The \gls{cu} is split further into the \gls{cp} and the \gls{up}. This logical split is performed via virtualization and softwarization, which allow different functions to be executed at different locations and on different platforms across the network. These functional units, shown in Fig.~\ref{fig:ORAN_arch}, %\ref{fig:E2interface},
%A robust discussion of the advantages of this disaggregation can be found in \cite{polese2022understanding}.  
% All the of components shown in Fig. \ref{fig:E2interface} 
can be abstracted from the physical infrastructure and deployed as software components (e.g., microservices, containers). %Large scale virtualization allows rapid scaling of compute resources and promises long term optimization and power saving. 
This architecture enables a decoupling between hardware and software components, sharing of hardware among different tenants, portability and execution on general-purpose hardware, and automated deployment of RAN functions. It is essential to grasp how virtualization and security considerations impact resource utilization, as this understanding is critical for designing solutions that can accommodate the security overhead effectively without overloading resources and introducing excessive latency.
   
\noindent $\bullet$~\textbf{\gls{ric}:} O-RAN introduces two \glspl{ric}, the non-RT \gls{ric} and the near-RT \gls{ric}. The \glspl{ric} are both designed to enable intelligent decision-making, network monitoring, and control via \gls{ai} and \gls{ml} solutions that are fed with data transferred across O-RAN open interfaces. The near-RT \gls{ric} receives data over the E2 interface and handles control loops on a time scale between 10 ms and 1 s, using plug-and-play components called xApps. The non-RT \gls{ric} collects data via the O1 interface and operates at time scales higher than 1 second using rApps, and is embedded in the \gls{smo} framework.
% The main function of the \gls{ric} is to use Key Performance Metric (KPM) data and leverage ML algorithms to determine and apply control policies and actions. While there are two RICs in the O-RAN framework, the non-RT \gls{ric} and the near-RT \gls{ric}, 
Although both \glspl{ric} play a vital role in the lifecycle management of O-RAN systems, in this paper, we focus on the near-RT \gls{ric} only. The near-RT \gls{ric} is the most sensitive to latency and overhead introduced by security mechanisms due to its stringent operational time scale and its tight coordination with controlled \glspl{gnb}. Indeed, as illustrated in Fig.~\ref{fig:ORAN_arch}%\ref{fig:E2interface}
, the near-RT \gls{ric} interfaces with the \glspl{cu} and \glspl{du} of the distributed \glspl{gnb}. On the contrary, since the non-RT \gls{ric} operates with latency timescales of 1s or greater, the cost of securing it (in terms of overhead and latency) is incremental and marginal.
% The near-RT \gls{ric} also hosts xApps that can be used to perform radio resource management \cite{polese2022understanding}.

\noindent $\bullet$~\textbf{Open Interfaces:} While decoupling hardware and software creates an open environment for faster development, it also introduces the need for interoperable interfaces. These are some of the key elements necessary to overcome the traditional RAN black-box approach as they expose network parameters to the \glspl{ric} and enable data analytics and ML-enabled control. 
The O1 interface is the primary interface with the \gls{smo} and the non-RT \gls{ric} and is responsible for enabling operations and maintenance. Similarly, the O2 interface connects the \gls{smo} to the O-Cloud, the abstraction for the infrastructure supporting O-RAN virtualization. The A1 interface connects the two \glspl{ric} and is used for deploying policy-based guidance. The E2 interface is the key interface that connects the near-RT \gls{ric} to the RAN (see Fig.~\ref{fig:ORAN_arch}). The E2 interface enables the collection of metrics from the RAN to the near-RT \gls{ric} and allows the \gls{ric} to control multiple functions in the disaggregated \gls{gnb}. Finally, the Open Fronthaul connects a \gls{du} to one or multiple \glspl{ru} inside the same \gls{gnb}~\cite{OranWG4}. The Open Fronthaul is further broken down into four distinct planes: (C)ontrol, (U)ser, (S)ynchronization, and (M)anagement plane. For example, the \gls{up} carries the actual user data in the form of \gls{iq} samples, while the S-plane carries timing and synchronization messages. The requirements in terms of both traffic type and security vary greatly between the planes. 
A comprehensive discussion of these and additional interfaces can be found in \cite{polese2023understanding_official, abdalla2022toward, upadhyaya2022prototyping, ericcson, OranWG4, OranWG3}. It is imperative to secure such interfaces as they might transport sensitive user data and network telemetry. The \textit{O-RAN ALLIANCE WG11: Security Work Group} requires specific security functions to be supported on each interface along with what protocol will provide those functions \cite{OranWG11-secreqspec, OranWG11}. Additional interface details are specified by \textit{WG4: Open Fronthaul Interfaces Workgroup}~\cite{OranWG4,OranWG4-M} and \textit{WG5: Open F1/W1/E1/X2/Xn Interface Workgroup}~\cite{OranWG5}. Table \ref{tab:interface req} lists each interface, security function, and protocol as specified by the O-RAN ALLIANCE.

% \begin{table}
% \centering
%     \begin{tabular}{|c|c;{1pt/1pt}l;{1pt/1pt}l;{1pt/1pt}l|c|c|} \hline
%     \multirow{2}{*}{\textbf{Interface}} & \multicolumn{4}{c|}{\textbf{Functions}} & \multirow{2}{*}{\textbf{Protocols}} & \textbf{Overhead} \\
%      & \multicolumn{1}{l;{1pt/1pt}} C & I & A & R &  & (Bytes) \\ \hline\hline
%     A1, 02 & \checkmark & \checkmark & \checkmark & \checkmark & TLS & 25 - 40\\ \hline
%     O1 & \checkmark & \checkmark & \checkmark &  & TLS & 25 - 40\\ \hline
%     F1-C & \checkmark & \checkmark &  & \checkmark & TLS & 25 - 40\\ \hline
%     E2 & \checkmark & \checkmark & \checkmark & \checkmark & IPsec & 57 - 76\\ \hline
%     F1/W1/E1/X2/Xn & \checkmark & \checkmark & \checkmark &  & IPsec & 57 - 76\\ \hline
%     \begin{tabular}[c]{@{}c@{}}Open Fronthaul \\C,M Plane\end{tabular} &  &  & \checkmark &  & 802.1X & NA \\ \hline
%     \begin{tabular}[c]{@{}c@{}}Open Fronthaul \\S,U Plane\end{tabular} &  &  & &  & none & NA \\ \hline
%     \end{tabular}
%     \caption{O-RAN Interface security functions (C:~Confidentiality; I:~Integrity: A:~Authentication; R:~Replay protection) and protocols required by O-RAN ALLIANCE WGs 11 and 5. }
%     \label{tab:interface req}
% \end{table}

\begin{table}[htb]
\centering
\resizebox{\linewidth}{!}{%
\begin{tabular}{|c|c|c|c|c|c|c|} \hline
Interface & \multicolumn{4}{c|}{Function} & Protocol & Overhead  \\
          & C & I & A & R                 &          & (Bytes)   \\ \hline\hline
A1, O2    & \checkmark & \checkmark & \checkmark & \checkmark & TLS & $\ge 25$ \\ \hline
O1        & \checkmark & \checkmark & \checkmark & & TLS & $\ge 25$ \\ \hline
F1- C     & \checkmark & \checkmark &   & \checkmark & TLS & $\ge 25$ \\ \hline
E2        & \checkmark & \checkmark & \checkmark & \checkmark & IPsec & $\ge 57$ \\ \hline
F1, W1, E1, X2, Xn & \checkmark & \checkmark & \checkmark & & IPsec & $\ge 57$ \\ \hline
Fronthaul- M & \checkmark & \checkmark & \checkmark & & SSHv2 or TLS & $\ge 28$ or $\ge 25$ \\ \hline
Fronthaul- C,U,S     &   &   & \checkmark  &                   & 802.1x   & N/A        \\ \hline
\end{tabular}
}
\caption{O-RAN Interface security functions (C:~Confidentiality; I:~Integrity: A:~Authentication; R:~Replay protection) and protocols specified by O-RAN ALLIANCE WGs 4, 5 \& 11. }
\label{tab:interface req}
\end{table}

\subsection{Network Security Protocols} \label{sec:back:security}

There are many useful models to describe the set of network security services. Here we focus on the four functions used by \textit{O-RAN ALLIANCE WG 11}: Confidentiality, Integrity, Authentication, and Replay Protection. Confidentiality ensures the payload of the message cannot be read while the data is in transit. Integrity guarantees the content of the message is not changed in transit. Authentication, in this context, guarantees the endpoints of a conversation are who they say they are. Finally, Replay Protection ensures a pre-recorded message cannot be sent as a new message in the future.

There are a wide range of protocols used to provide these network security functions at various layers of the network protocol stack. In this paper, we consider: \gls{tls}, which operates at the transport layer, \gls{ipsec}, which operates at the network layer, and \gls{macsec}, which operates at the data link layer. While each protocol can provide similar general security functions, there are also distinctions among them. %Here, we present a brief overview of these three protocols.

\gls{tls}~\cite{tls1.3} is a widely used transport layer security protocol that provides authentication during the initial handshake process, and offers confidentiality by using a suite of algorithms to encrypt the transport layer payload. \gls{tls} also provides integrity and replay protection through the use of a message authentication code to prevent replay attacks. Newer releases of \gls{tls} (e.g., \gls{tls} 1.3) have faster \gls{sa} establishment when compared to previous versions of \gls{tls} and \gls{ipsec}. After the initial \gls{sa} is established, \gls{tls} adds about 25-40 Bytes of overhead to each packet.

\gls{ipsec} works at the network layer and offers several modes of operation. The primary security modes are \gls{ah}~\cite{ipsec_ah}, which provides integrity, authentication, and replay attack protection; and \gls{esp}~\cite{ipsec_esp}, which additionally provides  encryption. In practice, \gls{esp} is almost exclusively used and the O-RAN ALLIANCE has mandated its use. \gls{ipsec} can also operate in either transport or tunnel mode. Tunnel mode creates a new \gls{ip} header for each packet and protects the integrity of both the data and the original \gls{ip} header 
% for each packet
\cite{frankel2005guide}. On the contrary, transport mode only secures the network layer payload. O-RAN specifications mandate the support of tunnel mode, while support for transport mode is optional. Even though \gls{ipsec} uses a slightly different handshake to establish \glspl{sa} compared to \gls{tls}, it still offers all the same security functions and utilizes the same cryptographic functions as \gls{tls} after the \gls{sa} is established. When using \gls{esp} and tunnel mode, \gls{ipsec} adds at least 57 Bytes of overhead to each packet. However, in practice this is often higher because the cryptographic functions are block ciphers, requiring a fixed-size input. 

\gls{macsec}~\cite{macsec}, is used for securing point-to-point connections at the data link layer. \gls{macsec} offers two modes: encryption on or off. Both modes provide integrity, replay protection, and authentication, but only the encryption on mode offers confidentiality. Unlike \gls{tls} and \gls{ipsec}, the \gls{macsec} standard specifies a single cryptographic algorithm, AES128-GCM, though implementations using AES256-GCM are available. \gls{macsec} adds a fixed 32-byte header to all packets. However, similar to the other encryption protocols, \gls{macsec} uses AES for confidentiality, which is a block cipher. Therefore, when encryption is enabled the overhead may be larger than 32 Bytes. Another security protocol that operates at the data link layer is IEEE 802.1x~\cite{802.1x}, which provides a standard for port-based authentication on \glspl{lan}. 802.1x only provides periodic authentication (a common default is every 60 minutes) of devices connected over physical ports and no other security functions.

One of the primary differences between all these protocols is the way in which \glspl{sa} are established. However, this happens infrequently; for \gls{tls} and \gls{ipsec} by default the \glspl{sa} expire after 8 hours, and for \gls{macsec} the \gls{sa} expires after 1 hour. While this re-authentication time is configurable, the default values used in the majority of applications, on the order of a few hours, are considered secure. It should be noted that very frequent (on the order of seconds) re-authentication greatly reduces throughput. Additionally, all three protocols allow re-authentication before the current \gls{sa} expires, guaranteeing continuous operation. Because this establishment happens very infrequently, it has a negligible impact on overhead and resource utilization. For this reason, we do not deeply analyze the initial handshake or \gls{sa} establishment in this article. On the other hand, there are numerous differences in the encryption algorithms used and security services provided which, as we will show, have a significant impact on performance and resource utilization.  %and, for this reason, we analyze these differences and their inherent cost in detail.

\subsection{The Latency Cost of Security}\label{types_of_delay}
To properly evaluate how security impacts network performance, it is important first to understand how the different security mechanisms affect the way data flows through the network. Specifically, since security adds overhead (both in terms of data and procedures), we need to establish a model that captures the effect that security has on total delay. 
In packet-switched networks, there are four primary sources of delay at each node along the path: \emph{queuing} delay ($D_{que}$), \emph{propagation} delay ($D_{prop}$), \emph{transmission} delay ($D_{trans}$), and \emph{nodal processing} delay ($D_{proc}$) \cite{kurose1986computer}. The total delay can be expressed as
\begin{equation}
    D_{total}=D_{que}+D_{prop}+D_{trans}+D_{proc}.
    \label{eq:delay}
\end{equation}

\noindent $\bullet$~\textbf{Queuing Delay:}\label{queuing} This delay considers the duration that packets spend in the processing queue at each interface along the end-to-end path. In our test environment, there is almost no competing traffic, allowing us to safely assume \(D_{que}=0\). In more complex networks, this assumption may not hold. 

To analyze queuing delay further, we model the network nodes using an M/M/1 queue with packet arrivals forming a Poisson process. Then the average queuing delay is given by \(D_{que}=\frac{1}{\mu-\lambda} - \frac{1}{\mu}\) where \(\lambda\) is the traffic arrival rate and \(\mu\) is the queue service rate. The difference between unsecured or \gls{pt} and secured or \gls{ct} delays depends on the overhead added by encryption, denoted as \(\epsilon\). The value of \(\epsilon\) is protocol-specific, but typically falls within the range of 60 Bytes (480 bits) or less. The difference in queuing delay is given by \(\Delta D_{que} =\frac{1}{\mu-(\lambda+\epsilon)} - \frac{1}{\mu-\lambda} \). When \(\mu - \lambda \gg \epsilon\), the approximation \(\Delta D_{que} \approx 0\) is valid. Even in significantly more congested networks, the queuing delay remains largely unaffected by the presence or absence of encryption. To illustrate, in our network supporting \(\mu = 10\) Gbps speeds, this approximation holds (i.e., \(\Delta D_{que} \le 1\) 
 \textmu s) when the arrival rate \(\lambda \le 9.78 \) Gbps.

\noindent $\bullet$~\textbf{Propagation Delay:}\label{delay_prop} The propagation delay is strictly a function of the physical length and propagation speed of the link. The propagation speed depends on the link type but is typically on the order of \(2\times 10^8\) m/s \cite{kurose1986computer}. For our environment, we assume a length of 100m giving \(D_{prop}=0.05\) \textmu s. This will change for other systems but will remain constant regardless of encryption.

\noindent $\bullet$~\textbf{Transmission Delay:} The transmission delay is a function of the packet size (in bits), \(L\), and the link transmission rate, \(R\), which is defined as \(D_{trans}=L/R\). For any given system, \(R\) is fixed but \(L\) will increase with encryption. Table \ref{table:D_trans} lists the calculated transmission delays, with and without encryption, based on the average packet length of the three types of packets we observe and describe in detail in Sec.~\ref{securingE2}. 

\begin{table}[htb]
    \centering
    \begin{tabular}{| p{2cm} || c | c |}
         \hline
         Packet Type & Plain Text & Cypher Text\\
         \hline
         \hline
         SACK & 62 B : \(0.0496 \mu\)s & 138 B : \(0.1104 \mu\)s\\
         Short E2AP & 195 B : \(0.1560 \mu\)s & 255 B : \(0.2040 \mu\)s\\
         Long E2AP & 1425 B : \(1.140 \mu\)s & 1485 B : \(1.188 \mu\)s\\
         \hline
    \end{tabular}
    \caption{Calculated transmission delay for 3 types of packets with and without encryption.}
    \label{table:D_trans}
\end{table}

\noindent $\bullet$~\textbf{Processing Delay:} Typically, this is defined as the time required for intermediate nodes to inspect the packet header and determine the appropriate path for the packet, whether at layer 2 for switching or layer 3 for routing. This time can also encompass other factors, such as bit-level error-checking \cite{kurose1986computer}. It's important to note that while the fronthaul network is expected to function as a pure layer-2 network (i.e., no routing involved), the E2 interface can be deployed over a layer-3 network, necessitating routing at certain intermediate nodes.

In our analysis, we include the encryption delay within the processing delay, as the cryptographic functions are integral to passing the payload to lower or higher layers in the network stack. Notably, this encryption delay occurs only at the sending and receiving nodes. Intermediate nodes do not need to engage in cryptographic operations because the Ethernet and \gls{ip} headers are sent in \gls{pt}.

\section{O-RAN Security Evaluation Platforms}\label{s: environment}
% \rev{
% Should there be some sort of brief introduction here?
% %
% LB: probably, something brief like the following. Also, should the name of the section be more on the interface side, since this is what is discussed therein?

% In this section, we give an overview the O-RAN interface implementation that we used in this work. We detail the E2 interface implementation in Section~\ref{sec:e2}, and the Open Fronthaul interface implementation in Section~\ref{of_description}.
% }
%\rev{Can we also add a detailed figure of the platform/experimental setup? Let's call it a security blabla O-RAN blabla platform maybe?}
In this section, we describe the interface security evaluation platforms we used in this work. Specifically, the E2 interface implementation is illustrated in Sec.~\ref{sec:e2}, and the Open Fronthaul interface implementation is described in Sec.~\ref{of_description}. The results obtained on both platforms will be presented in Secs.~\ref{s:E2} and~\ref{s:OF}.

\subsection{E2 Interface}
\label{sec:e2}

We extensively use Colosseum~\cite{bonati2021colosseum}, the world's largest wireless network emulator with hardware in-the-loop, to evaluate the E2 interface. Colosseum supports experimental research through virtualized protocol stacks, enabling users to test full-protocol solutions at scale, with real hardware devices, in realistic emulated RF environments with complex channel interactions. The key building blocks for deploying full protocol stacks are 128 \glspl{srn}. Each \gls{srn} consists of a 48-core Intel Xeon E5-2650 CPU with an NVIDIA Tesla k40m GPU connected to a USRP X310 \gls{sdr}. Users can instantiate custom protocol stacks by deploying \glspl{lxc} on the bare-metal \glspl{srn}.

%The Colosseum environment also provides several RF scenarios created through its Massive Channel Emulator (MCHEM) and managed by simple APIs. This enables users to test custom protocol stacks in a wide range of both artificially constructed (such as a fixed path loss) and realistic RF scenarios built from field observations. 

%Significant work has been done to create an open source O-RAN environment on top of this publicly available experimental platform. 
We utilize the srsRAN-based SCOPE \cite{bonati2021scope} framework to implement a softwarized RAN for both the \gls{gnb} and multiple \glspl{ue}. SCOPE extends srsLTE (now srsRAN) version 20.04 by adding an E2 interface, several open APIs to facilitate run-time reconfiguration of the \gls{gnb}, and additional data collection tools. We utilize the ColO-RAN \cite{polese2022colo} framework for the near-RT \gls{ric} implementation. ColO-RAN offers a minimal version of the \gls{osc} near-RT \gls{ric} (Bronze release) tailored to execute on Colosseum via an \gls{lxc}-packaged set of Docker containers. 
% The ColO-RAN \gls{lxc} includes a minimal near-RT \gls{ric} in the form of several Docker containers to provide the functions of the near-RT \gls{ric}. 
In particular, it provides an E2 interface implementation compliant with O-RAN specifications that can be used to interface with the RAN nodes for data collection and control. ColO-RAN also offers an extensible xApp template that collects basic \glspl{kpm} from the \gls{gnb} and can send control actions to it.

To secure the E2 interface, we add the strongSwan~\cite{project-strongswan} open source \gls{ipsec}-based VPN to both the SCOPE and ColO-RAN \glspl{lxc}. The full \gls{ipsec} configuration is described in paragraph \ref{securingE2}. We also add several simple scripts to automate data collection on E2 interface performance which we make open source for public use and further research.

\begin{figure}[htb]
    \centering
    \includegraphics[width=\linewidth]{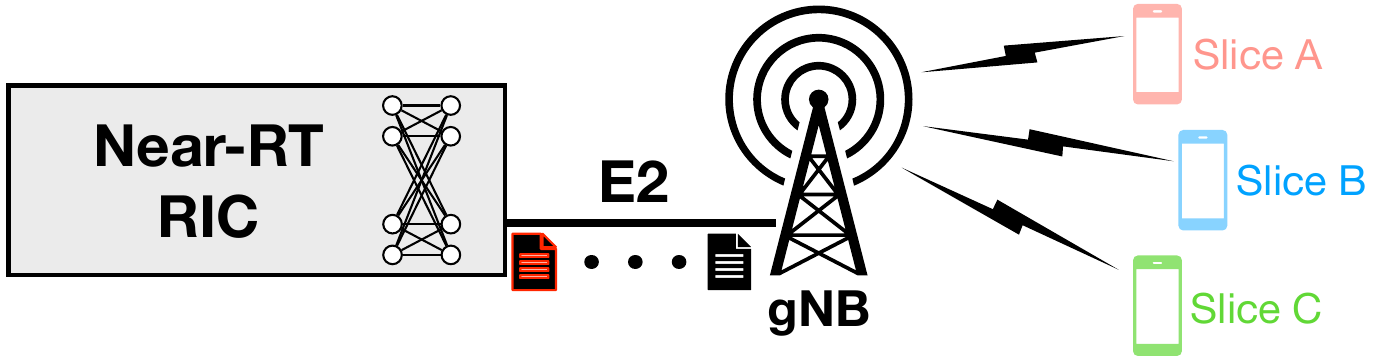}
    \caption{O-RAN testbed used to study security in the E2 interface, consisting of three \glspl{ue}, a \gls{gnb}, and the near-RT \gls{ric}. Each component is implemented in an \gls{lxc} on top of an \gls{srn} within Colosseum.}
    \label{fig:system}
\end{figure}

\subsubsection{E2 Experimental Setup}\label{securingE2}

Our experimental system (see Fig.~\ref{fig:system}) is composed of \textcolor{black}{up to twelve blocks: 10} \glspl{ue}, a \gls{gnb}, and the near-RT \gls{ric}. Each block is implemented through \gls{lxc} on separate \glspl{srn}. The \glspl{ue} are connected to the \gls{gnb} over an emulated RF channel where each \gls{ue} is assigned \textcolor{black}{to one of three} unique slices representing the three main use cases for 5G: enhanced Mobile Brodband (eMMB), Ultra Reliable Low Latency Communications (URLLC), and massive Machine Type Communications (mMTC) \cite{7784887}. Each slice has its own traffic pattern \textcolor{black}{generated from real world 5G traffic traces as described in \cite{groen2023tractor}. First, we use a variety of applications to generate traffic for each network slice. For eMBB, we stream videos, browse the Internet, and transfer large files. For URLLC, we conduct both voice phone calls, video chat, and utilize real time AR applications. For mMTC we capture texts and background traffic from all apps when the phone is not actively being used. This is not the typical example of mMTC traffic, such as IoT applications. However, it does fit nicely in the fundamental definition of mMTC because it is low throughput, latency tolerant communication from numerous applications. Next, we built a traffic generator tool to replay the traffic between the UE and gNB. The traffic generator emulates the original traffic by reading the length field for each packet and sending a random byte string of the appropriate length at the time indicated by the packet timestamp. This enables us to replicate the timing, length, and direction of all data sent between the UE and gNB, while completely anonymizing the actual payload within our experimental test bed. Our O-RAN test bed further emulates the channel conditions between the gNB and UE based on measured channel conditions for a real deployed cellular system. This allows us to accurately capture the O-RAN KPIs as if the original communication were taking place in a deployed O-RAN test bed.} The \gls{gnb} is connected to the near-RT \gls{ric} over a wired 10 Gbps backbone network. We implement the sample \gls{kpm} monitoring xApp from \cite{polese2022colo} that periodically polls the \gls{gnb} for \textcolor{black}{up to 31} \glspl{kpm} for each \gls{ue}. \textcolor{black}{This generates between 200 Kbps of traffic with 3 \glspl{ue} sending 6 \glspl{kpm} up to 3.5 Mbps of traffic with 10 \glspl{ue} sending 31 \glspl{kpm}} on the E2 interface. %The \glspl{kpm} we collected include: slice ID, downlink queue length, transmit rate in Mbps, ratio of granted to requsted PRBs, the number of PRBs assigned to the slice, and the number of transmitted packets in the last period. 

\begin{figure}[t]
    \centering
    \includegraphics[width=\linewidth]{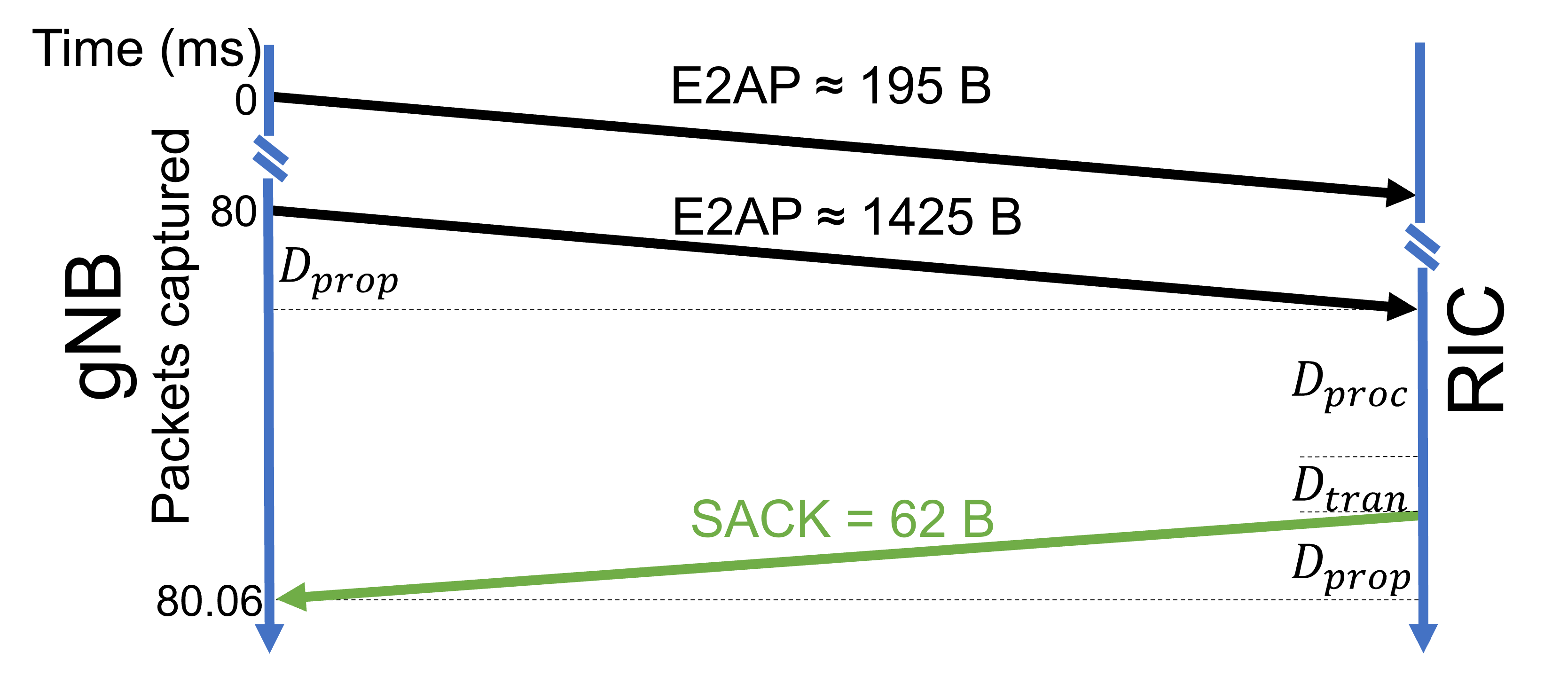}
    \caption{E2 traffic displays a common pattern of one small E2AP packet then one large E2AP packet followed by a \gls{sack} as seen in this flow graph.}
    \label{fig:flowgraph}
\end{figure}

\begin{figure}[tb]
    \centering
    \includegraphics[width=\linewidth]{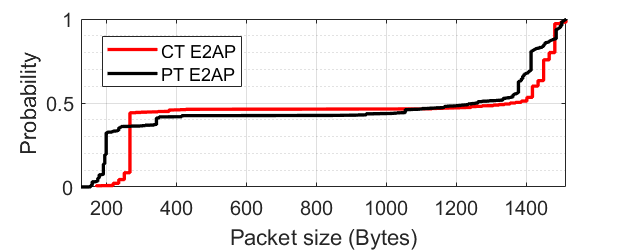}
    \caption{CDF of \gls{e2ap} packet length for both \gls{pt} and \gls{ct} traffic.}
    \label{fig:e2 packet size cdf}
\end{figure}

We capture all traffic traversing the E2 interface at the \gls{gnb} for over 20 minutes. The \gls{sctp} is used as the transport layer protocol for all traffic. Fig.~\ref{fig:flowgraph} illustrates a typical example of the captured traffic. First, the \gls{gnb} sends a small data packet followed by a large data packet using \gls{e2ap} over \gls{sctp}. 
Fig.~\ref{fig:e2 packet size cdf} shows the empirical CDF of the \gls{e2ap} packet sizes with and without encryption. %The average size for both short and long \gls{e2ap} packets measured on our testbed are reported in Table \ref{table:D_trans}.
%While this pattern is consistent, the exact size of both the small and large X2AP packets varies.
Finally, the near-RT \gls{ric} responds with a fixed-size \gls{sack} packet (62 Bytes). This pattern is consistent because \gls{sctp} specifies that a \gls{sack} should be generated for every second packet received \cite{stewart2007stream}. \textcolor{black}{Any additional processing delay introduced by IPsec to the E2 interface takes place at the kernel level, outside of the srsLTE software stack. Even though} we do not have access to the precise time when the \gls{gnb} starts processing or transmitting a packet\textcolor{black}{,} we can observe the delay between the transmission of the large \gls{e2ap} packet and the reception of the \gls{sack} at the \gls{gnb}. For these reasons, we first study the effect of encryption on the \gls{sack}. 

%While we are able to capture the elapsed time between packet captures, we do not have access to the precise time when the \gls{gnb} starts processing or transmitting a packet. However, we can observe the delay between the transmission of the large \gls{e2ap} packet and the reception of the \gls{sack} at the \gls{gnb}. For these reasons, we first study the effect of encryption on the \gls{sack}.

After establishing a baseline performance without encryption, we add O-RAN-compliant encryption as specified in \cite{OranWG3} to the E2 interface. We implement \gls{ipsec} with \gls{esp} in tunnel mode. For each packet, tunnel mode creates a new \gls{ip} header, and protects the integrity of both the data and original \gls{ip} header \cite{frankel2005guide}. We use AES256 for encryption and SHA2-256 for the authentication hash function. AES256 is a high-speed symmetric encryption algorithm that uses a fixed block size of 128 bits and a key size of 256 bits and performs 14 transformation rounds \cite{AES}. SHA2-256 uses eight 32-bit words and performs 64 transformation rounds to compute a 256-bit hash \cite{fips2012180}. However, only the first 128 bits of the hash are included in the \gls{ipsec} trailer. \gls{ipsec}, as configured in our test, provides all the required services listed in the O-RAN specifications for E2~\cite{OranWG3}. %: confidentiality, integrity, replay protection, and authentication. 
With this configuration, \gls{ipsec} adds at least 57 Bytes of overhead to each packet. However, because both AES256 and SHA2-256 require fixed input block sizes, padding may be added causing the overhead to further increase. For example, encrypting the \gls{sack} adds 76 Bytes for a total \gls{ct} \gls{sack} length of 138 Bytes. We generate the same \gls{ue} traffic described earlier, poll the \gls{gnb} for the same \glspl{kpm}, and again capture the traffic traversing the E2 interface at the \gls{gnb}.

\subsection{Open Fronthaul}\label{of_description}

\begin{figure}[htb]
    \centering
    \includegraphics[width=\linewidth]{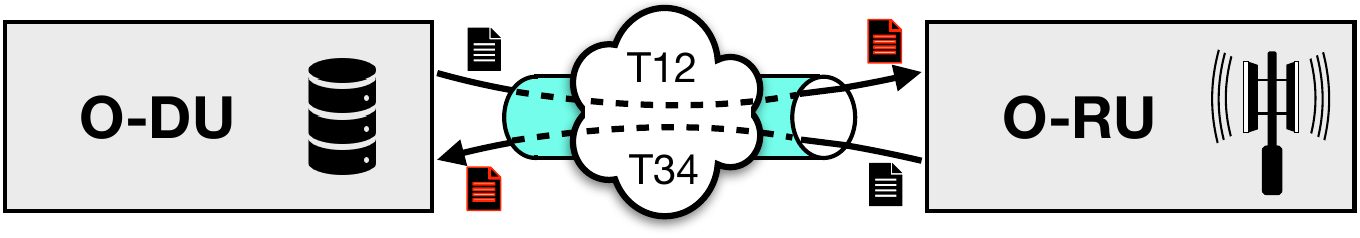}
    \caption{The Open Fronthaul (shown in blue) connects the O-DU and O-RU, potentially across a wider network (cloud). The O-RAN ALLIANCE WG4 specifies transmission delays for the downlink (T12) and the uplink (T34).}
    \label{fig:OF_system}
\end{figure}

An overview of the Open Fronthaul interface is shown in Fig.~\ref{fig:OF_system}. While the Open Fronthaul interface can be viewed as a single connection (blue pipe) passing through other networks (white cloud), the O-RAN ALLIANCE WG4 specifies the allowable latency's for downlink (T12) and uplink (T34) separately.

We leverage the NVIDIA \gls{arc} platform~\cite{aerialsdk} to capture traffic from a production-ready O-RAN and 5G-compliant system based on a 3GPP 7.2 split~\cite{OranWG4}. NVIDIA \gls{arc} combines the open-source project \gls{oai}~\cite{oai} for the higher layers of the protocol stack with the NVIDIA Aerial physical implementation, which runs on \gls{gpu} for inline acceleration (i.e., NVIDIA A100). The \gls{gpu} in an \gls{arc} server is combined with a programmable \gls{nic} (in our case, a Mellanox ConnectX-6 Dx) through \gls{rdma}, bypassing the CPU to transfer packets from the \gls{nic} to the \gls{gpu} itself. This makes it possible to implement a high-speed \gls{du}-side termination of the Open Fronthaul interface, capable of block floating point compression to ensure prompt delivery of the \gls{iq} samples and control messages to the \gls{ru}.

Specifically, we leverage a private 5G network deployed at Northeastern University (part of the X-Mili project)~\cite{villa2023x5g}, including 8 \gls{arc} nodes with a dedicated \gls{cn} and fronthaul infrastructure. The fronthaul infrastructure features a Dell S5248F-ON switch, with a Qulsar QG-2 acting as a grandmaster clock. The latter distributes \gls{ptp} and SyncE synchronization to the \gls{du} and the \gls{ru}. In our setup, the \gls{ru} is a Foxconn 4T4R unit operating in the $3.7-3.8$\:GHz band, and we use \gls{cots} 5G \glspl{ue} from OnePlus (AC Nord 2003)~\cite{oneplus}.

While the code base for the \gls{du} is open and potentially extensible to embed \gls{macsec}, the \gls{ru} comes with a closed-source FPGA-based termination for the fronthaul interface. This prevents us from directly enabling \gls{macsec} in our Open Fronthaul environment. Therefore, we adopt a trace-based approach and configure a port of the fronthaul switch to mirror the fronthaul traffic to a server running a packet capture. We built an emulation environment in our lab using two desktop computers with Intel i9-13900K CPUs with NVIDIA Mellanox ConnectX-4 Lx \glspl{nic} directly connected over a $10$\:Gbps Ethernet link.
%
% We consider two test benches to emulate the Open Fronthaul by replaying the captured traffic. 
% \begin{itemize}[wide, labelindent=0pt]
%     \item \textbf{Colosseum environment:} The first system utilizes two separate \glspl{srn} in Colosseum, equipped with $10$\:Gbps \glspl{nic} connected to the networking backbone of Colosseum (implemented through a BigSwitch \gls{sdn} Ethernet system, which we cannot control or configure).
%     \item \textbf{Home-grown environment:} The second environment is built in our lab using two desktop computers with Intel i9-13900K CPUs with NVIDIA Mellanox ConnectX-4 Lx NICs directly connected over a $10$\:Gbps Ethernet link. 
% \end{itemize}
% 
%In both systems, 
We leverage a Python script to re-play the original pcap file captured on the real Open Fronthaul. In this way, we can properly emulate the original Open Fronthaul capture and observe the impact of adding any encryption. The Linux kernel natively supports \gls{macsec}. We use the Ubuntu commands provided in documentation~\cite{Dubroca_2019} to configure \gls{macsec}. %The full \gls{macsec} configuration is described further in subsection~\ref{securing open fronthaul}. 

\subsubsection{Open Fronthaul Experimental Setup}\label{securing open fronthaul}

\begin{figure}[tb]
    \centering
    \includegraphics[width=\linewidth]{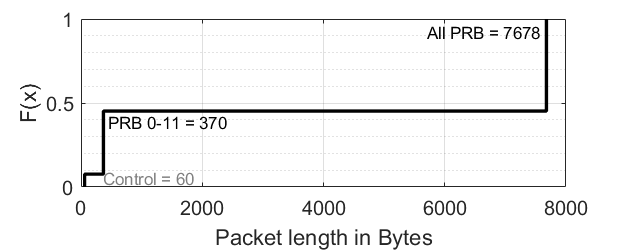}
    \caption{CDF of Open Fronthaul packet length, highlighting the three types of packets observed: \gls{cp}, \gls{up} with PRB 0-11, and \gls{up} with all PRBs.}
    \label{fig:of packet size cdf}
\end{figure}

We capture all the traffic traversing our Open Fronthaul described in Sec.~\ref{of_description} for over 20 minutes for several different traffic loads. In our environment, the Open Fronthaul uses the eCPRI protocol to send both \gls{cp} and \gls{up} messages. There are two network options to use eCPRI; the first uses the full network stack (UDP over \gls{ip}), while the second is designed for point-to-point networks and only uses the MAC layer. Our system is the second; our traffic only contains an Ethernet header with the eCPRI header and payload. While eCPRI supports payloads of up to 8192 Bytes, we consistently observe a maximum frame length of 7678 Bytes on the wire as shown in Fig.~\ref{fig:of packet size cdf}. 

Currently, the O-RAN ALLIANCE specifications do not call for any security on the Open Fronthaul (C, U, S) planes. However, there are known vulnerabilities to not securing the Open Fronthaul~\cite{10.1145/3495243.3558259, 10056720}. For this reason, there is a growing momentum for advocating to secure this vital link using \gls{macsec}~\cite{10.1145/3465481.3470080, 9604996}. \gls{macsec} can be configured with or without encryption and we test both modes of operation. As mentioned above, due to hardware limitations our experimental environment does not support \gls{macsec} hardware offloading in the \gls{nic} directly, so all \gls{macsec} operations are performed in software. %We observe similar patterns in results from both of our emulation environments described in Sec.~\ref{of_description}. All results shown here are from the local lab environment unless otherwise noted.

\section{E2 Experimental Results}\label{s:E2}
%In this section, we report the key results from the E2 Interface. 

\subsection{\gls{sack} Analysis}\label{SACK_analysis}

\begin{figure}[htb]
    \centering
    \includegraphics[width=\linewidth]{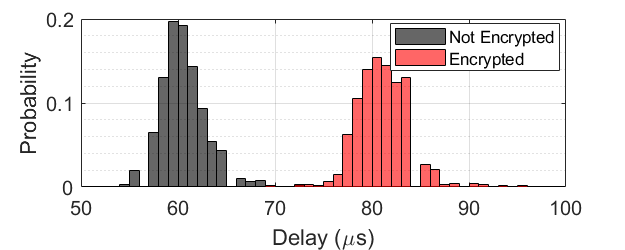}
    \caption{\gls{sack} delay distribution with and without encryption.}
    \label{fig:SACK}
\end{figure}

Fig. \ref{fig:SACK} shows the distribution of delay times for both \gls{pt} and \gls{ct} \glspl{sack}. It is immediately clear that encryption adds approximately $22$ \textmu s of delay on average. However, it is essential for O-RAN researchers, engineers, and system architects to fully understand both the cause and the impact of any extra overhead.

The \gls{sack} packet is ideal for an initial analysis because the packet size is fixed (PT=62 B, CT=138 B) and the total delay we observe in our experiments is very tightly grouped, as seen in Fig. \ref{fig:SACK}. The total average delay we observe for the \gls{pt} \gls{sack} is 61.12 \textmu s, while the \gls{ct} \gls{sack} is 82.64 \textmu s. From Fig. \ref{fig:flowgraph}, we can see that the total delay for a \gls{sack} can be expressed as \(D_{total}^{SACK} = 2\times D_{prop} + D_{trans} + D_{proc} \). Given the total delay, the transmission delays in Table~\ref{table:D_trans}, and the calculated propagation delay (Sec.~\ref{delay_prop}), we can calculate the average processing delay; for the \gls{pt} \gls{sack} it is 60.97 \textmu s while the \gls{ct} \gls{sack} is 82.64 \textmu s. In Colosseum, encryption on the E2 link adds approximately 22 \textmu s of delay to small packets. However, the near-RT \gls{ric} is designed to operate on scales from 10 ms to 1 s. \textbf{These results show that \textcolor{black}{for the E2 Interface}, with low traffic load (\(\leq 200kbps\)) and small packets (\(\leq 138 B\)), encryption has no meaningful impact on E2 interface traffic.}

\subsection{Packet Size Analysis}\label{ss:e2 packet size analysis}

Our initial results examining the \gls{sack} over the E2 interface (Sec.~\ref{SACK_analysis}) show that the processing delay accounts for at least \(99.75\%\) of the total delay for the 62 Byte \gls{sack}, regardless of encryption. In other words, the delay for short packets is dominated by the processing delay. Given the calculated propagation and transmission delays, it is likely that processing delay dominates in equation (\ref{eq:delay}) for larger packets as well. However, since it is not possible to know the exact start times for transmitting the long X2AP packets based on the traces (and we do not have any \gls{sack} for the Open Fronthaul) we conduct another experiment to confirm this hypothesis.

To fully quantify the effect of encryption for packets of various lengths, we use ping (ICMP echo) of various lengths to accurately capture the network \gls{rtt}. We start with a small payload for the ping and increment the size regularly. For each step size, we send 100 pings with 250 ms between each ping. Because there are no competing flows for this experiment the queuing delay is zero and the \gls{rtt} is expressed as
\begin{equation}
    RTT=2\times (D_{proc}+ D_{trans}+ D_{prop}).
        \label{eq:rtt}
\end{equation}

\begin{figure}[tb]
    \centering
    \includegraphics[width=\linewidth]{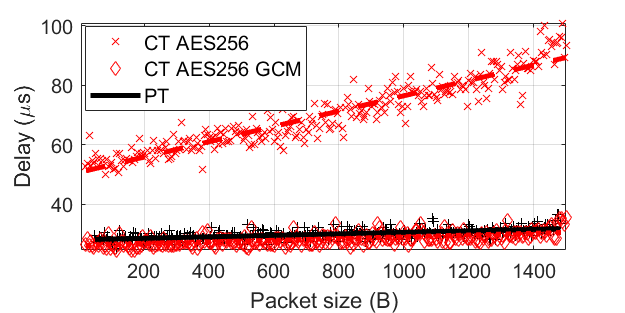}
    \caption{Processing delay as a function of packet size for \gls{pt} and \gls{ct} traffic over the E2 Interface.}
    \label{fig:process_delay}
\end{figure}

Given the calculated propagation and transmission delays, \(2\times D_{prop} = 0.1\) \textmu s, \(2\times D_{trans} \leq 2.4\) \textmu s, and the observed \(RTT\ge 50\) \textmu s, we can approximate equation \eqref{eq:rtt} as \(D_{proc} \approx \frac{RTT}{2}\). From the results shown in Fig.~\ref{fig:process_delay}, we observe that the processing delay increases with packet length when sending encrypted traffic using AES256. However, the processing delay difference between \gls{ct} and \gls{pt} is \(\Delta D_{proc} \leq 50\) \textmu s  for all tested packet sizes. In contrast, when we use AES256-GCM we see no difference in processing delay between the \gls{pt} and \gls{ct} traffic. We can conclude that, regardless of the encryption algorithm used, \textbf{for all packet sizes from 62 B to 1500 B, encryption has minimal impact on E2 traffic}.

\subsection{Throughput Analysis}\label{sss:throughput}

We also design an experiment to quantify the effect of encryption on the total traffic throughput, \(T\). %First, we found the maximum throughput of the system using iperf3. We found that the plaintext throughput is approximately 9.5 Gbps while the cyphertext throughput is approximately 500 Mbps. 
We use iperf3 to generate traffic at specific bit rates for 10 seconds and regularly increment the transmission rate. We record the actual throughput reported by the receiving node. We also capture CPU utilization on the \gls{gnb} (sending node) for each attempted transmission rate using iperf3. 

\begin{figure}[tb]
    \centering
    \includegraphics[width=\linewidth]{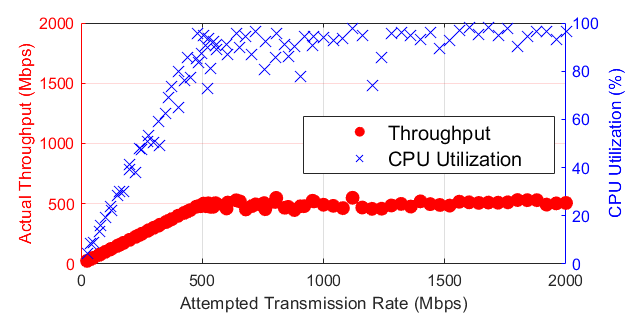}
    \caption{Measured throughput and CPU utilization (in blue) as a function of attempted transmission rate for \gls{ct} traffic (\gls{ipsec}) over the E2 Interface while using AES256-CBC.}
    \label{fig:ipsec throughput cpu}
\end{figure}

From Fig.~\ref{fig:ipsec throughput cpu} it can be seen that the maximum encryption rate our system is capable of when using AES256-CBC as the encryption algorithm is $\leq 575$ Mbps. Fig.~\ref{fig:ipsec throughput cpu} also shows the CPU utilization while using \gls{ipsec}. While CPU utilization does increase for both types of traffic, we can see that encryption is very CPU-intensive. When sending \gls{pt}, the CPU utilization increases proportional to \(0.00365 \times T \)~\cite{groenCost}, whereas Fig.~\ref{fig:ipsec throughput cpu} shows the CPU utilization for \gls{ct} grows proportional to \(0.2 \times T\) until it reaches saturation. Therefore, we conclude that encrypting all traffic increases CPU utilization by roughly two orders of magnitude compared to \gls{pt} traffic. In our system, using AES256 \textbf{CPU utilization is a limiting factor for encrypting traffic} when the attempted transmission rate is $\geq 575$ Mbps. 

We ran the same experiment with a wide variety of encryption algorithms and key sizes. We observed the same pattern where the \gls{ct} throughput increases linearly until it plateaus. The maximum throughput varies greatly based on the encryption algorithm used, as seen in Table~\ref{table: max throughput}. \textbf{The specific algorithm implementation is the most significant factor \textcolor{black}{when implementing IPsec on the E2 interface}} with AES-GCM vastly outperforming the other algorithm implementations. Galois/Counter Mode (GCM) simultaneously provides confidentiality (using counter mode) and authentication (using arithmetic in the Galois field GF($2^{n}$)), where $n$ is the key size \cite{mcgrew2004galois}. These operations can be performed in parallel, offering greater performance than other modes such as CBC, which require chaining of operations. 

\begin{table}[htb]
\centering
\renewcommand*{\arraystretch}{1.2}
\begin{tabular}{|c| c|}
     \hline
     \textbf{Encryption algorithm} & \textbf{Throughput}\\
     \hline
     \hline
     AES128-CBC & $505$\:Mbps\\
     \hline
     AES256-CBC & $512$\:Mbps\\
     \hline
     AES128-CCM & $573$\:Mbps\\
     \hline
     AES256-CCM & $573$\:Mbps\\
     \hline
     ChaCha20-Poly1305 & $989$\:Mbps\\
     %\hline
     %AES256-GCM64 & $1370$\:Mbps\\
     \hline
     AES256-GCM  & $1370$\:Mbps \\
     \hline
\end{tabular}
\caption{30 second throughput for various encryption algorithms. The algorithm and implementation have a greater impact on system performance than key size, with AES-GCM performing the best.}
\label{table: max throughput}
\end{table}

One key observation is that for all implementations there was virtually no difference in performance based on key size (AES128 versus AES256). In addition to the difference in key length, AES128 only uses 10 transformation rounds while AES256 uses 14 rounds~\cite{AES}. The additional rounds for AES256 consist of the operations: SubBytes, ShiftRows, and MixColumns which are highly optimized for performance and do not add any significant delay. We encourage all system designers to use the longer 256-bit key length as it provides higher security with virtually no impact on performance. While individual systems and encryption algorithms may be different, these experiments show that CPU utilization is a key trade-off.

\section{Open Fronthaul Experimental Results}\label{s:OF}

\subsection{Packet Size Analysis}\label{ss: OF packet size}

\begin{figure}[tb]
    \centering
    \includegraphics[width=\linewidth]{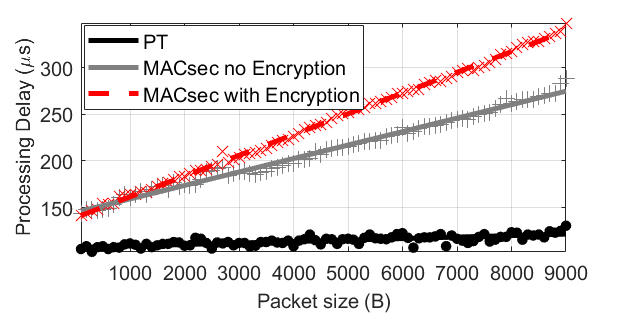}
    \caption{Processing delay as a function of packet size over the Open Fronthaul.}
    \label{fig: MACSEC packet size delay}
\end{figure}

We repeat the experiment described in Sec.~\ref{ss:e2 packet size analysis} over the Open Fronthaul interface and plot the processing delay without and with \gls{macsec} in Fig.~\ref{fig: MACSEC packet size delay}. It is immediately clear that using \gls{macsec} on the Open Fronthaul interface increases delay and has a greater impact on larger packets than smaller packets. Using \gls{macsec}, with or without encryption, increases the delay for small packets by $\approx 39$ \textmu s. For the maximum packet size of 9000 Bytes, the processing delay due to \gls{macsec} with no encryption increases to $\approx 153$ \textmu s. However, the cost of using encryption increases faster than that for \gls{macsec} without encryption, reaching a total increase of $\approx 218$ \textmu s. 

\begin{table}[h]
\centering
\resizebox{\linewidth}{!}{%
\begin{tabular}{|r||c|c|c|c|c|c|c|c|c|c|c|c|c|c|} \hline
\multicolumn{1}{|c||}{O-RU} & \multicolumn{14}{c|}{{\cellcolor[rgb]{0.902,0.902,0.902}}O-DU Category} \\
\multicolumn{1}{|c||}{Cat} & \multicolumn{1}{c}{{\cellcolor[rgb]{0.902,0.902,0.902}}A} & \multicolumn{1}{c}{{\cellcolor[rgb]{0.902,0.902,0.902}}B} & \multicolumn{1}{c}{{\cellcolor[rgb]{0.902,0.902,0.902}}C} & \multicolumn{1}{c}{{\cellcolor[rgb]{0.902,0.902,0.902}}D} & \multicolumn{1}{c}{{\cellcolor[rgb]{0.902,0.902,0.902}}E} & \multicolumn{1}{c}{{\cellcolor[rgb]{0.902,0.902,0.902}}F} & \multicolumn{1}{c}{{\cellcolor[rgb]{0.902,0.902,0.902}}G} & \multicolumn{1}{c}{{\cellcolor[rgb]{0.902,0.902,0.902}}H} & \multicolumn{1}{c}{{\cellcolor[rgb]{0.902,0.902,0.902}}I} & \multicolumn{1}{c}{{\cellcolor[rgb]{0.902,0.902,0.902}}J} & \multicolumn{1}{c}{{\cellcolor[rgb]{0.902,0.902,0.902}}K} & \multicolumn{1}{c}{{\cellcolor[rgb]{0.902,0.902,0.902}}L} & \multicolumn{1}{c}{{\cellcolor[rgb]{0.902,0.902,0.902}}M} & {\cellcolor[rgb]{0.902,0.902,0.902}}N \\ \hhline{|~|:==============|}
O & {\cellcolor[rgb]{1,0.447,0.463}}3000 & {\cellcolor[rgb]{1,0.447,0.463}}399 & {\cellcolor[rgb]{1,0.447,0.463}}379 & {\cellcolor[rgb]{1,0.447,0.463}}359 & {\cellcolor[rgb]{1,0.447,0.463}}339 & {\cellcolor[rgb]{0.75,0.75,0.75}}319 & {\cellcolor[rgb]{0.75,0.75,0.75}}299 & {\cellcolor[rgb]{0.75,0.75,0.75}}279 & {\cellcolor[rgb]{0.678,0.847,0.902}}259 & {\cellcolor[rgb]{0.678,0.847,0.902}}239 & {\cellcolor[rgb]{0.678,0.847,0.902}}219 & {\cellcolor[rgb]{0.678,0.847,0.902}}199 & {\cellcolor[rgb]{0.678,0.847,0.902}}179 & {\cellcolor[rgb]{0.678,0.847,0.902}}159 \\ \hhline{|~||--------------|}
P & {\cellcolor[rgb]{1,0.447,0.463}}2949 & {\cellcolor[rgb]{1,0.447,0.463}}348 & {\cellcolor[rgb]{0.75,0.75,0.75}}328 & {\cellcolor[rgb]{0.75,0.75,0.75}}308 & {\cellcolor[rgb]{0.75,0.75,0.75}}288 & {\cellcolor[rgb]{0.678,0.847,0.902}}268 & {\cellcolor[rgb]{0.678,0.847,0.902}}248 & {\cellcolor[rgb]{0.678,0.847,0.902}}228 & {\cellcolor[rgb]{0.678,0.847,0.902}}208 & {\cellcolor[rgb]{0.678,0.847,0.902}}188 & {\cellcolor[rgb]{0.678,0.847,0.902}}168 & {\cellcolor[rgb]{0.678,0.847,0.902}}148 & {\cellcolor[rgb]{0.678,0.847,0.902}}128 & 108 \\ \hhline{|~||--------------|}
Q & {\cellcolor[rgb]{1,0.447,0.463}}2929 & {\cellcolor[rgb]{0.75,0.75,0.75}}328 & {\cellcolor[rgb]{0.75,0.75,0.75}}308 & {\cellcolor[rgb]{0.75,0.75,0.75}}288 & {\cellcolor[rgb]{0.678,0.847,0.902}}268 & {\cellcolor[rgb]{0.678,0.847,0.902}}248 & {\cellcolor[rgb]{0.678,0.847,0.902}}228 & {\cellcolor[rgb]{0.678,0.847,0.902}}208 & {\cellcolor[rgb]{0.678,0.847,0.902}}188 & {\cellcolor[rgb]{0.678,0.847,0.902}}168 & {\cellcolor[rgb]{0.678,0.847,0.902}}148 & {\cellcolor[rgb]{0.678,0.847,0.902}}128 & 108 & 88 \\ \hhline{|~||--------------|}
R & {\cellcolor[rgb]{1,0.447,0.463}}2909 & {\cellcolor[rgb]{0.75,0.75,0.75}}308 & {\cellcolor[rgb]{0.75,0.75,0.75}}288 & {\cellcolor[rgb]{0.678,0.847,0.902}}268 & {\cellcolor[rgb]{0.678,0.847,0.902}}248 & {\cellcolor[rgb]{0.678,0.847,0.902}}228 & {\cellcolor[rgb]{0.678,0.847,0.902}}208 & {\cellcolor[rgb]{0.678,0.847,0.902}}188 & {\cellcolor[rgb]{0.678,0.847,0.902}}168 & {\cellcolor[rgb]{0.678,0.847,0.902}}148 & {\cellcolor[rgb]{0.678,0.847,0.902}}128 & 108 & 88 & 68 \\ \hhline{|~||--------------|}
S & {\cellcolor[rgb]{1,0.447,0.463}}2889 & {\cellcolor[rgb]{0.75,0.75,0.75}}288 & {\cellcolor[rgb]{0.678,0.847,0.902}}268 & {\cellcolor[rgb]{0.678,0.847,0.902}}248 & {\cellcolor[rgb]{0.678,0.847,0.902}}228 & {\cellcolor[rgb]{0.678,0.847,0.902}}208 & {\cellcolor[rgb]{0.678,0.847,0.902}}188 & {\cellcolor[rgb]{0.678,0.847,0.902}}168 & {\cellcolor[rgb]{0.678,0.847,0.902}}148 & {\cellcolor[rgb]{0.678,0.847,0.902}}128 & 108 & 88 & 68 & 48 \\ \hhline{|~||--------------|}
T & {\cellcolor[rgb]{1,0.447,0.463}}2869 & {\cellcolor[rgb]{0.678,0.847,0.902}}268 & {\cellcolor[rgb]{0.678,0.847,0.902}}248 & {\cellcolor[rgb]{0.678,0.847,0.902}}228 & {\cellcolor[rgb]{0.678,0.847,0.902}}208 & {\cellcolor[rgb]{0.678,0.847,0.902}}188 & {\cellcolor[rgb]{0.678,0.847,0.902}}168 & {\cellcolor[rgb]{0.678,0.847,0.902}}148 & {\cellcolor[rgb]{0.678,0.847,0.902}}128 & 108 & 88 & 68 & 48 & 28 \\ \hhline{|~||--------------|}
U & {\cellcolor[rgb]{1,0.447,0.463}}2849 & {\cellcolor[rgb]{0.678,0.847,0.902}}248 & {\cellcolor[rgb]{0.678,0.847,0.902}}228 & {\cellcolor[rgb]{0.678,0.847,0.902}}208 & {\cellcolor[rgb]{0.678,0.847,0.902}}188 & {\cellcolor[rgb]{0.678,0.847,0.902}}168 & {\cellcolor[rgb]{0.678,0.847,0.902}}148 & {\cellcolor[rgb]{0.678,0.847,0.902}}128 & 108 & 88 & 68 & 48 & 28 & 8 \\ \hhline{|~||--------------|}
V & {\cellcolor[rgb]{1,0.447,0.463}}2829 & {\cellcolor[rgb]{0.678,0.847,0.902}}228 & {\cellcolor[rgb]{0.678,0.847,0.902}}208 & {\cellcolor[rgb]{0.678,0.847,0.902}}188 & {\cellcolor[rgb]{0.678,0.847,0.902}}168 & {\cellcolor[rgb]{0.678,0.847,0.902}}148 & {\cellcolor[rgb]{0.678,0.847,0.902}}128 & 108 & 88 & 68 & 48 & 28 & 8 & 0 \\ \hhline{|~||--------------|}
W & {\cellcolor[rgb]{1,0.447,0.463}}2809 & {\cellcolor[rgb]{0.678,0.847,0.902}}208 & {\cellcolor[rgb]{0.678,0.847,0.902}}188 & {\cellcolor[rgb]{0.678,0.847,0.902}}168 & {\cellcolor[rgb]{0.678,0.847,0.902}}148 & {\cellcolor[rgb]{0.678,0.847,0.902}}128 & 108 & 88 & 68 & 48 & 28 & 8 & 0 & 0 \\ \hhline{|~||--------------|}
X & {\cellcolor[rgb]{1,0.447,0.463}}2789 & {\cellcolor[rgb]{0.678,0.847,0.902}}188 & {\cellcolor[rgb]{0.678,0.847,0.902}}168 & {\cellcolor[rgb]{0.678,0.847,0.902}}148 & {\cellcolor[rgb]{0.678,0.847,0.902}}128 & 108 & 88 & 68 & 48 & 28 & 8 & 0 & 0 & 0 \\ \hhline{|~||--------------|}
Y & {\cellcolor[rgb]{1,0.447,0.463}}2769 & {\cellcolor[rgb]{0.678,0.847,0.902}}168 & {\cellcolor[rgb]{0.678,0.847,0.902}}148 & {\cellcolor[rgb]{0.678,0.847,0.902}}128 & 108 & 88 & 68 & 48 & 28 & 8 & 0 & 0 & 0 & 0 \\ \hhline{|~||--------------|}
Z & {\cellcolor[rgb]{1,0.447,0.463}}2749 & {\cellcolor[rgb]{0.678,0.847,0.902}}148 & {\cellcolor[rgb]{0.678,0.847,0.902}}128 & 108 & 88 & 68 & 48 & 28 & 8 & 0 & 0 & 0 & 0 & 0 \\ \hline
\end{tabular}
}
\caption{The O-RAN ALLIANCE WG4 specified maximum transmission delays in \textmu s. For our system the \gls{ru}/\gls{du} combinations in the red region support \gls{macsec} with encryption, the grey region support \gls{macsec} without encryption, and the blue region is supported without any \gls{macsec}.}
\label{table: OF max latency}
\end{table}

Considering the latency requirements discussed in Sec.~\ref{of_description} for the Open Fronthaul and the observed large frame sizes, it is evident that, for certain systems, \textbf{the use of \gls{macsec} may significantly impact Open Fronthaul traffic}. The O-RAN ALLIANCE specifies the minimum and maximum latency supported by different equipment combinations for \gls{ru} and \gls{du}, as detailed in \cite{OranWG4} and reproduced in Table~\ref{table: OF max latency}. For any system, the chart will exhibit three distinct regions, contingent on the configuration and use of \gls{macsec}. The red region of the chart represents \gls{ru}/\gls{du} combinations that support \gls{macsec} using encryption in the system we used to evaluate performance. The grey region, signifying \gls{macsec} without encryption, broadens the range of \gls{ru}/\gls{du} combinations at our disposal. Finally, the blue region introduces a substantial number of additional combinations that the considered evaluation system can accommodate without using \gls{macsec} altogether.

It is important for researchers and system engineers to understand the \gls{ru} and \gls{du} capabilities and requirements and the cost of enabling \gls{macsec} with or without encryption before deploying systems. This work demonstrates that there are combinations of \gls{ru}/\gls{du} that support \gls{macsec}, though the number of combinations is significantly smaller than with no security. Therefore, security impacts the size of the feasibility region of certain O-RAN deployments, and it is fundamental to understand the security-latency trade-off prior to deployment.

% \begin{figure}[tb]
%     \centering
%     \includegraphics[width=.95\linewidth]{OpenInterface/max_latency_annotated.png}
%     \caption{The O-RAN ALLIANCE WG4~\cite{OranWG4} specified maximum transmission delays for the downlink (T12) and the uplink (T34) in $\mu$s. The \gls{ru}/\gls{du} combinations in the red region support \gls{macsec} with encryption, those in the black region support \gls{macsec} without encryption, and those in the green region are supported without any \gls{macsec}.\rev{how about white, blue, and light blue regions? Is this a screenshot? If yes, then I suggest creating a new one.}}
%     \label{fig: OF max latency}
% \end{figure}

\subsection{Throughput Analysis}\label{ss: OF throughput analysis}
\begin{figure}[tb]
    \centering
    \includegraphics[width=\linewidth]{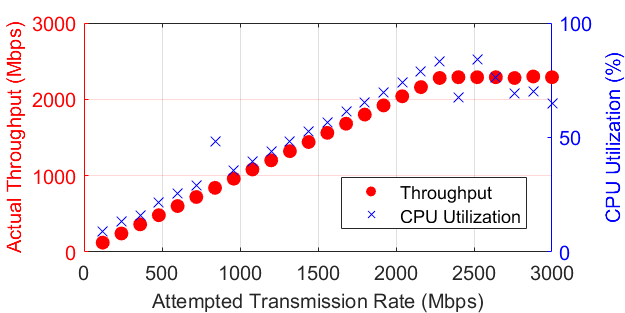}
    \caption{Actual \gls{ct} throughput (red) and CPU utilization (blue) for different attempted transmission rates using \gls{macsec}.}
    \label{fig: MACSEC throughput cpu}
\end{figure}

We repeat the same experiment described in Sec.~\ref{sss:throughput} for the Open Fronthaul. The maximum throughput for our local emulation environment is approximately 2300 Mbps, as seen in Fig.~\ref{fig: MACSEC throughput cpu}. %We observe similar patterns in our Colosseum environment but with a maximum throughput of approximately 1200 Mbps. 
Based on these results we again conclude that \textbf{CPU utilization is a key factor when adding \gls{macsec}.} However, unlike in the case of the E2, the CPU utilization is not above 95\%. This suggests there may be other limiting factors, which we discuss further in Sec.~\ref{ss: io}. 

We also study the impact of total throughput on delay specifically for the Open Fronthaul. For this experiment, we again use iperf3 to generate increasing traffic load. Simultaneously, we use ping with two different fixed-size packets to evaluate the \gls{rtt}. We repeat the experiment with small (300 Byte) and large (8172 Byte) packets. We subtract the known transmission and propagation delays. In this experiment, unlike the previous ones, there may be queuing delays at the \gls{nic} as the iperf3 and ICMP traffic compete for the same physical interface. The \gls{rtt} is therefore given by $RTT=2\times (D_{proc}+ D_{queue})$. However, in Sec.~\ref{queuing} we show that the added queuing delay is negligible when the queue arrival rate is less than 9.78 Gbps. Thus, we can again approximate $RTT\approx 2\times D_{proc}$ and solve for the added processing delay.

\begin{figure}[tb]
    \centering
    \includegraphics[width=\linewidth]{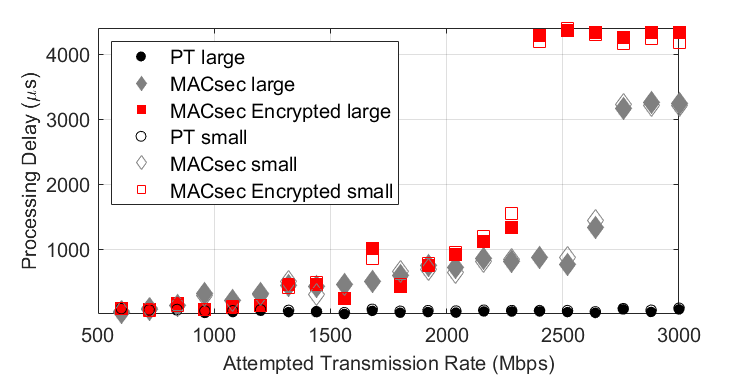}
    \caption{Open Fronthaul processing delay as a function of attempted transmission rate.}
    \label{fig: MACSEC throughput delay}
\end{figure}

Fig.~\ref{fig: MACSEC throughput delay} shows the results of this experiment and illustrates the impact of throughput on delay. First, we observe that packet size tends to have less impact than the overall transmission rate. Second, we again observe that \gls{macsec} with encryption has a higher cost, in terms of increased latency, compared to \gls{macsec} without encryption. While both methods achieve a similar maximum throughput, the maximum latency for \gls{macsec} with encryption is about 4.3 ms while the maximum latency for \gls{macsec} without encryption is about 3.2 ms. In other words, \textbf{the specific \gls{macsec} configuration is a key factor in determining \textcolor{black}{Open Fronthaul} throughput and latency.}

\subsection{\gls{mtu} Analysis}

The \gls{mtu} of a network path is determined by the minimum \gls{mtu} supported by any device along that path. This becomes critical when analyzing the Open Fronthaul. Not only could a disaggregated \gls{ru} and \gls{du} be deployed in physically separate locations, but a single entity may not control the entire network path between them. %This became apparent between our two emulation environments. We noticed significant differences in performance between the Colosseum and local lab environment. Upon further examination, we discovered that intermediate switches in Colosseum have a maximum \gls{mtu} of 1500. This initial result pointed toward the impact \gls{mtu} size can have.

We designed an experiment to better understand the impact of different \gls{mtu} sizes on the Open Fronthaul traffic. We send a large file, 4.66 GB, using iperf3 over our emulated Open Fronthaul interface using \gls{macsec} with encryption enabled. We chose a large enough file to saturate the link for around 15 seconds based on the maximum observed throughput of approximately 2.5 Gbps. We repeat this experiment while incrementing the \gls{mtu} size from 1400 Bytes to 9000 Bytes. We record the total time to complete the transmission as well as the average throughput for the duration of the transmission. 

\begin{figure}[tb]
    \centering
    \includegraphics[width=\linewidth]{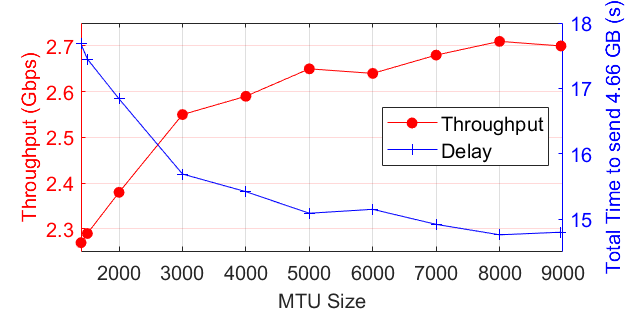}
    \caption{\gls{mtu} size impacts throughput (red) and delay (blue) in our deployed system using \gls{macsec} with encryption. A larger \gls{mtu} provides better performance.}
    \label{fig:MTU}
\end{figure}

Fig.~\ref{fig:MTU} shows the results of our \gls{mtu} analysis. As the \gls{mtu} increases, the time to send a large file decreases and the effective throughput increases. Increasing the \gls{mtu} size from 1400 Bytes to 9000 Bytes improves performance (reduces delay) by approximately 20\%.  From these results, we conclude that, for the large packet sizes and high data rates observed in the Open Fronthaul, \textbf{\gls{mtu} size is a critical factor.}

% \begin{figure}[htb]
%      \centering
%      \begin{subfigure}{\linewidth}
%          \centering
%          \includegraphics[width=.95\linewidth]{PT_CPU_1.png}
%          \caption{\gls{pt} CPU utilization grows proportional to \(0.00365 \times T\).}
%          \label{fig:PT_CPU}
%      \end{subfigure}
%      \hfill
%      \begin{subfigure}{\linewidth}
%          \centering
%          \includegraphics[width=.95\linewidth]{CT_CPU_1.PNG}
%          \caption{\gls{ct} CPU utilization grows proportional to \(0.2 \times T\) until it reaches the maximum utilization allowable. After this threshold, throughput and CPU utilization remain constant.}
%          \label{fig:CT_CPU}
%      \end{subfigure}
%      \hfill
%      \caption{Actual throughput and CPU utilization as functions of attempted transmission rate for \gls{pt} and \gls{ct} traffic.}
%      \label{fig:CPU}
% \end{figure}

\section{Cost of Security Key Principles} \label{s:analysis}
Although the results presented so far show trends and phenomena that are transferable to any deployment configuration, it is important to point out that the exact values of all these results depend heavily on the specific system resources and network topology considered. In the following, we make an effort to extend our analysis to a more general configuration. Specifically, we present four key principles from lessons that we learned to enable researchers and system architectures to build security by design in O-RAN systems.

%While the E2 interface standards call for the use of \gls{ipsec}, other open interfaces require the use of TLS 1.2, though use of TLS 1.3 is recommended. There are some differences in the way \gls{ipsec} and TLS protocols initially establish a connection or SA, however, we are confident we can extend the results generally without loss of accuracy because \gls{ipsec} and TLS use the same underlying encryption and hashing algorithms once the SA is established \cite{groen2023implementing}. TLS 1.3 has the fastest SA establishment, making it hardest to confidently extrapolate the results directly from \gls{ipsec} to TLS 1.3, but the general trends in performance should still hold. Regardless of the security protocol used, system designers should only use modern secure encryption algorithms.

\subsection{Sufficient Compute Resources}

Sufficient processing power is one of the key trade-offs required to enable encryption. Generalizing Table~\ref{table:D_trans}, we can conclude that for any system operating over Gigabit Ethernet, or faster, \(\Delta D_{trans} << \Delta D_{proc}\). From the calculations in Sec.~\ref{delay_prop}, we can see that for any network where the total distance between the base station and near-RT \gls{ric} is in the order of tens of kilometers or less, \(\Delta D_{prop} << \Delta D_{proc}\). We are confident that any additional queuing delays due to encryption, \(\Delta D_{que} \approx 0 \) for nearly all conditions as shown in Sec.~\ref{queuing}. Therefore, for most O-RAN systems, the total delay \emph{cost of encryption}, \(\Delta D_{total}\) can be approximated as $\Delta D_{total}\approx \Delta D_{proc}$. 
%the key lessons learned we present in this section represent a valuable resource for system designers and architects. The four key considerations are: Adequate Computing Resources, Specific Encryption Algorithms, I/O Bottlenecks, and \gls{mtu} Size.

Even if these assumptions are not universally applicable to a specific O-RAN system, adding encryption will impact the processing delay. Any disaggregated \gls{gnb} component must have enough CPU resources to manage all of its explicit functions. While 3GPP standards allow for the use of secure gateways to deal with encryption, O-RAN standards do not. This means that as the total traffic over the E2 and Open Fronthaul interface (and other encrypted interfaces) increases, the CPU resources needed for encryption alone will increase. This could potentially impact the scalability and costs of \gls{gnb} components, especially the DU. System designers can choose to add dedicated hardware for the encryption to offload the CPU burden, increase the total compute resources, or set strict limits on the amount of traffic that can be sent over O-RAN open interface. Therefore, \textbf{system engineers must understand the amount of traffic expected across a given interface and include the overhead of encryption for that level of traffic in their compute and hardware acceleration budget}.

\subsection{Specific Encryption Algorithms}

The fundamental protection provided by \gls{ipsec} is through the encryption algorithm used to secure the payload. \textcolor{black}{A complete discussion of all possible encryption algorithms (for example} StrongSwan, supports 49 different encryption algorithms~\cite{strongswanIKEv2Cipher}) \textcolor{black}{is beyond the scope of this paper}. Some of these only perform encryption and require a separate integrity hashing algorithm, while others are Authenticated Encryption with Associated Data (AEAD) algorithms that do not require separate integrity hashing algorithms. Even limiting the scope of algorithms to those that are currently known to be secure, there is a huge array of options. \textcolor{black}{However,} as seen in Fig.~\ref{fig:process_delay}, choosing the specific encryption algorithm is incredibly important.

%The specific encryption algorithm or method of protection is vital. The fundamental protection provided by \gls{ipsec} is through the encryption algorithm used to secure the payload. One popular open source \gls{ipsec} configuration tool, StrongSwan, supports 49 different encryption algorithms~\cite{strongswanIKEv2Cipher}. Some of these only perform encryption and require a separate integrity hashing algorithm, while others are Authenticated Encryption with Associated Data (AEAD) algorithms that do not require separate integrity hashing algorithms. Even limiting the scope of algorithms to those that are currently known to be secure, there is a huge array of options. As seen in Fig.~\ref{fig:process_delay}, choosing the specific encryption algorithm is incredibly important. 

In contrast, \gls{macsec} uses a standard encryption algorithm but offers two modes: plain text payload or encrypted payload. As seen in Sec.~\ref{ss: OF packet size} and~\ref{ss: OF throughput analysis} this distinction makes a significant difference in the processing delay. System engineers should make careful decisions about which method to use for what traffic. For example, the \gls{up} traffic is already encrypted by the PDCP layer between the \gls{ue} and \gls{cu}. In other words, the \gls{up} traffic crossing the Open Fronthaul already has confidentiality. Using \gls{macsec} without encryption for \gls{up} traffic will provide the other necessary security functions (authentication, integrity, and replay protection) at a lower cost. On the other hand, the synchronization and management plane may not provide confidentiality at other layers, requiring the use of \gls{macsec} with encryption. 

While the O-RAN ALLIANCE guidance requires similar security functions to be provided for most interfaces, \textbf{the specific security protocol and encryption algorithm used must be carefully chosen to meet both security and system performance requirements.}

\subsection{I/O Bottlenecks}\label{ss: io}

In Fig.~\ref{fig: MACSEC throughput cpu} we see that the actual throughput reaches a plateau while the CPU utilization is only around 80\%. Also, as discussed in Sec.~\ref{sss:throughput}, while we observe similar patterns with all encryption algorithms, for some the CPU utilization plateaus around only 50\%.  While this observation does not diminish the importance of optimizing the specific protocol and encryption algorithm used, it does indicate another factor plays a key role; namely the I/O operations between kernel space and user space in Linux.

Network packet processing for encrypted traffic in the Linux kernel can be significantly slow due to context switching associated with system calls and transitional copy operations in packet traversal through all network layers \cite{ullah2020ipsec}. Specifically for StrongSwan, there is a significant bottleneck from user/kernel space context switching. Prior work \cite{ullah2020ipsec} achieves roughly a 3.5x increase in throughput and a 2.5x decrease in latency by removing or optimizing these I/O operations. Other methods of improving I/O bottlenecks include offloading workload from the CPU to the \gls{nic}, reducing the number of interrupts generated by incoming traffic, optimizing the basic Linux Kernel network stack, and moving network functions entirely to user space such as with DPDK \cite{7110118}. \textbf{It is vital for O-RAN system designers, working with distributed and compute constrained devices, to understand the existing network stack I/O bottlenecks and properly optimize each component.}

These first three principles are often highly dependent on each other. For example, our Open Fronthaul system using the Mellanox ConnectX-6 does not support hardware acceleration of \gls{macsec}. However, it does support hardware acceleration of \gls{ipsec} when using AES-GCM. It also supports \gls{tls} data-path offloading and keeping all the \gls{tls} functions in the \gls{up}. A comprehensive systems engineering approach is essential for optimizing the interplay among these considerations. 

\subsection{\gls{mtu} size}

We clearly see that a larger \gls{mtu} provides better performance in our emulation environments. Much of this improvement comes from the reduction in overhead by using less total packets. For example, with a payload of exactly 8000 Bytes, the amount of overhead with an \gls{mtu} size of 4000 Bytes will be double the overhead compared to a network with an \gls{mtu} of 8000 Bytes. The experiments we perform involve transferring a large, fixed amount of data. While this is a good approximation to the Open Fronthaul and provides measurable metrics, it does not exactly match the nature of the Open Fronthaul interface. Instead, we observe that the Open Fronthaul sends fixed packet sizes (7678 Bytes) at regular intervals. 

\begin{figure}[tb]
    \centering
    \includegraphics[width=\linewidth]{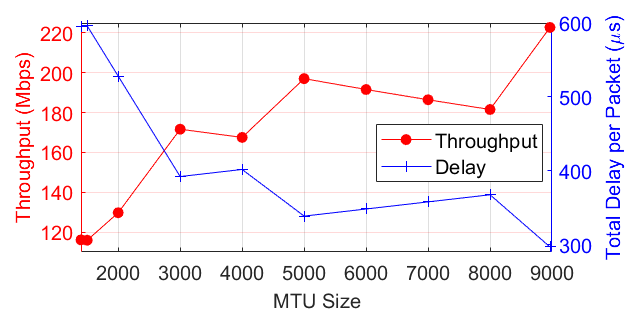}
    \caption{Modeled impact of \gls{mtu} size on throughput (red) and delay (blue) using \gls{macsec} with encryption. A larger \gls{mtu} provides better performance.}
    \label{fig:MTU model}
\end{figure}

We built a model based on our observations to understand the impact of \gls{mtu} size on regular, fixed-size packets. We use 8192 Bytes as the payload size because it is the maximum eCPRI payload size. We calculate the total overhead based on the number of packets the payload must be fragmented into and calculate the delay of each packet based on packet size using Eq.~\ref{eq:delay} and Fig.~\ref{fig: MACSEC packet size delay}. We calculate the throughput as the total bits sent over the total delay. The results of the model are seen in Fig.~\ref{fig:MTU model}.

The general trends of our model closely match our experimental results. However, there is one key difference. Our experiment shows continuous improvement in performance with increasing \gls{mtu} size. For a fixed file size, this intuitively makes sense as it minimizes the total amount of overhead compared to the total file size. In contrast, our model shows that setting the \gls{mtu} to 5000 Bytes gives better performance than 8000 Bytes. In other words, for fixed size packets transmitted with regular frequency, if fragmentation must happen, it is best to split the payload as evenly as possible. This is because the processing delay increases significantly with packet size, as shown in Fig.~\ref{fig: MACSEC packet size delay}, while the other delays are essentially constant. Splitting a fixed amount of data into two equal chunks results in lower processing delay than one large and one small chunk.

In either case, the best option is to ensure the entire network path has an \gls{mtu} of 9000 Bytes. However, it is likely that a single entity will not control the entire network path between the \gls{ru} and \gls{du} in future O-RAN systems. In this case, \textbf{system engineers must determine the maximum \gls{mtu} of the network path and carefully select the optimal \gls{mtu} size for the distributed \gls{ru} and \gls{du}.}

\section{Conclusions} \label{s:conclusion}

5G stands as a critical strategic technology, offering enhanced performance and data-driven intelligent capabilities \cite{DoD}. O-RAN, driven by its open interfaces and adaptable \glspl{ric}, plays a pivotal role in harnessing these capabilities. It empowers the customization of radio resource management through powerful \gls{ml}-driven xApps/rApps while exposing valuable telemetry. Given the paramount importance of safeguarding O-RAN's open links, this article conducts a comprehensive experimental and theoretical analysis of the E2 and Open Fronthaul interfaces, aligning with O-RAN specifications \cite{OranWG3, OranWG11, OranWG11-secreqspec, OranWG4, OranWG5}, using the world's largest emulator, Colosseum \cite{bonati2021colosseum} and a private, 5G and O-RAN compliant network.

We implement various security protocols (\gls{ipsec} for E2, \gls{macsec} for Open Fronthaul) and employ diverse encryption techniques %, such as AES-CBC, AES-GCM, and ChaCha20, 
to assess their impact on critical network performance metrics. Our analysis covers key parameters like processing delay and throughput, accompanied by detailed quantitative insights. Notably, we find that the cost of securing O-RAN for the E2 interface remains low, while \gls{macsec} is more likely to exert a significant impact on the Open Fronthaul interface. % due to its latency demands and increased processing delay. 
We conclude that system designers must ensure O-RAN disaggregated nodes possess sufficient computing resources, make judicious protocol and encryption algorithm selections, optimize I/O bottlenecks, and manage local \gls{mtu} settings with an understanding of the end-to-end network \gls{mtu}.

Although significant progress has been made, there is still substantial work to fully comprehend O-RAN security and the associated costs. %Hidden expenses may arise when securing other interfaces or facilitating secure xApp development and deployment. 
Ensuring security in a multi-vendor xApp environment is a largely open problem with possible hidden expenses. One potential solution is adopting a zero-trust approach~\cite{ramezanpour2022intelligent, DoD}, which presents a promising framework but has yet to be implemented.
%demands further implementation to understand the Cost of Securing O-RAN. 
Continued efforts are required to shape the future of secure O-RAN systems.

% \section*{Acknowledgment}

% \rev{Add something here if needed.}

\footnotesize  % for natbib
\bibliographystyle{IEEEtran}
\bibliography{IEEEabrv,references}

\begin{thebibliography}{10}
\providecommand{\url}[1]{#1}
\csname url@samestyle\endcsname
\providecommand{\newblock}{\relax}
\providecommand{\bibinfo}[2]{#2}
\providecommand{\BIBentrySTDinterwordspacing}{\spaceskip=0pt\relax}
\providecommand{\BIBentryALTinterwordstretchfactor}{4}
\providecommand{\BIBentryALTinterwordspacing}{\spaceskip=\fontdimen2\font plus
\BIBentryALTinterwordstretchfactor\fontdimen3\font minus
  \fontdimen4\font\relax}
\providecommand{\BIBforeignlanguage}[2]{{%
\expandafter\ifx\csname l@#1\endcsname\relax
\typeout{** WARNING: IEEEtran.bst: No hyphenation pattern has been}%
\typeout{** loaded for the language `#1'. Using the pattern for}%
\typeout{** the default language instead.}%
\else
\language=\csname l@#1\endcsname
\fi
#2}}
\providecommand{\BIBdecl}{\relax}
\BIBdecl

\bibitem{polese2023understanding_official}
M.~Polese, L.~Bonati, S.~D’oro, S.~Basagni, and T.~Melodia, ``{Understanding
  O-RAN: Architecture, Interfaces, Algorithms, Security, and Research
  Challenges},'' \emph{IEEE Communications Surveys \& Tutorials}, 2023.

\bibitem{OranWG1}
{O-RAN Working Group 1}, ``{O-RAN} {A}rchitecture {D}escription 5.00,''
  ORAN.WG1.O-RAN-Architecture-Description-v05.00, Tech. Rep., July 2021.

\bibitem{thaliath2022predictive}
J.~Thaliath, S.~Niknam, S.~Singh, R.~Banerji, N.~Saxena, H.~S. Dhillon, J.~H.
  Reed, A.~K. Bashir, A.~Bhat, and A.~Roy, ``{Predictive Closed-Loop Service
  Automation in O-RAN Based Network Slicing},'' \emph{IEEE Communications
  Standards Magazine}, vol.~6, no.~3, pp. 8--14, Sep. 2022.

\bibitem{shen2022security}
C.~Shen, Y.~Xiao, Y.~Ma, J.~Chen, C.-M. Chiang, S.~Chen, and Y.~Pan, ``Security
  {T}hreat {A}nalysis and {T}reatment {S}trategy for {ORAN},'' in \emph{2022
  24th International Conference on Advanced Communication Technology
  (ICACT)}.\hskip 1em plus 0.5em minus 0.4em\relax IEEE, 2022, pp. 417--422.

\bibitem{ramezanpour2022intelligent}
K.~Ramezanpour and J.~Jagannath, ``Intelligent zero trust architecture for
  5{G}/6{G} networks: {P}rinciples, challenges, and the role of machine
  learning in the context of {O}-{RAN},'' \emph{Computer Networks}, p. 109358,
  2022.

\bibitem{abdalla2022toward}
A.~S. Abdalla, P.~S. Upadhyaya, V.~K. Shah, and V.~Marojevic, ``Toward {N}ext
  {G}eneration {O}pen {R}adio {A}ccess {N}etworks--{W}hat {O-RAN} {C}an and
  {C}annot {D}o!'' \emph{IEEE Network}, 2022.

\bibitem{groen2023implementing}
J.~Groen, S.~D'Oro, U.~Demir, L.~Bonati, M.~Polese, T.~Melodia, and
  K.~Chowdhury, ``{Implementing and Evaluating Security in O-RAN: Interfaces,
  Intelligence, and Platforms},'' \emph{arXiv preprint arXiv:2304.11125}, 2023.

\bibitem{10.1145/3465481.3470080}
J.~Y. Cho and A.~Sergeev, ``{Secure Open Fronthaul Interface for 5G
  Networks},'' in \emph{Proceedings of the 16th International Conference on
  Availability, Reliability and Security}, ser. ARES 21.\hskip 1em plus 0.5em
  minus 0.4em\relax New York, NY, USA: Association for Computing Machinery,
  2021.

\bibitem{10056720}
W.~Tiberti, E.~Di~Fina, A.~Marotta, and D.~Cassioli, ``{Impact of
  Man-in-the-Middle Attacks to the O-RAN Inter-Controllers Interface},'' in
  \emph{2022 IEEE Future Networks World Forum (FNWF)}, 2022, pp. 367--372.

\bibitem{10.1145/3495243.3558259}
S.-H. Liao, C.-W. Lin, F.~A. Bimo, and R.-G. Cheng, ``{Development of C-Plane
  DoS Attacker for O-RAN FHI},'' in \emph{Proceedings of the 28th Annual
  International Conference on Mobile Computing And Networking}, ser. MobiCom
  '22, 2022, p. 850–852.

\bibitem{ericcson}
J.~Boswell and S.~Poretsky, ``Security considerations of {O}pen {RAN},''
  \emph{Stockholm: Ericsson}, 2020.

\bibitem{OranWG3}
{O-RAN Working Group 3}, ``Near-{R}eal-time {RAN} {I}ntelligent {C}ontroller
  {A}rchitecture \& {E2} {G}eneral {A}spects and {P}rinciples,''
  ORAN.WG3.E2GAP-v02.02, Tech. Rep., July 2022.

\bibitem{DoD}
\BIBentryALTinterwordspacing
D.~of~Defense, ``5{G} {S}trategy {I}mplementation {P}lan,'' Department of
  Defense, Tech. Rep., December 2020, accessed: 2024-02-15. [Online].
  Available: \url{https://apps.dtic.mil/sti/pdfs/AD1118833.pdf}
\BIBentrySTDinterwordspacing

\bibitem{9604996}
D.~Dik and M.~S. Berger, ``Transport security considerations for the open-ran
  fronthaul,'' in \emph{2021 IEEE 4th 5G World Forum (5GWF)}, 2021, pp.
  253--258.

\bibitem{bonati2021colosseum}
L.~Bonati, P.~Johari, M.~Polese, S.~D’Oro, S.~Mohanti, M.~Tehrani-Moayyed,
  D.~Villa, S.~Shrivastava, C.~Tassie, K.~Yoder \emph{et~al.}, ``Colosseum:
  {L}arge-scale wireless experimentation through hardware-in-the-loop network
  emulation,'' in \emph{2021 IEEE International Symposium on Dynamic Spectrum
  Access Networks (DySPAN)}.\hskip 1em plus 0.5em minus 0.4em\relax IEEE, 2021,
  pp. 105--113.

\bibitem{bonati2022openran}
L.~Bonati, M.~Polese, S.~D’Oro, S.~Basagni, and T.~Melodia, ``Open{RAN}
  {G}ym: {A}n {O}pen {T}oolbox for {D}ata {C}ollection and {E}xperimentation
  with {AI} in {O-RAN},'' in \emph{2022 IEEE Wireless Communications and
  Networking Conference (WCNC)}.\hskip 1em plus 0.5em minus 0.4em\relax IEEE,
  2022, pp. 518--523.

\bibitem{bonati2021scope}
L.~Bonati, S.~D'Oro, S.~Basagni, and T.~Melodia, ``{SCOPE}: An open and
  softwarized prototyping platform for {N}ext{G} systems,'' in
  \emph{Proceedings of the 19th Annual International Conference on Mobile
  Systems, Applications, and Services}, 2021, pp. 415--426.

\bibitem{polese2022colo}
M.~Polese, L.~Bonati, S.~D’Oro, S.~Basagni, and T.~Melodia, ``Col{O-RAN}:
  {D}eveloping machine learning-based x{A}pps for open {RAN} closed-loop
  control on programmable experimental platforms,'' \emph{IEEE Transactions on
  Mobile Computing}, 2022.

\bibitem{villa2023x5g}
D.~Villa, I.~Khan, F.~Kaltenberger, N.~Hedberg, R.~S. da~Silva, A.~Kelkar,
  C.~Dick, S.~Basagni, J.~M. Jornet, T.~Melodia, M.~Polese, and
  D.~Koutsonikolas, ``{An Open, Programmable, Multi-vendor 5G O-RAN Testbed
  with NVIDIA ARC and OpenAirInterface},'' 2023.

\bibitem{groenCost}
J.~Groen, B.~Kim, and K.~Chowdhury, ``{The Cost of Securing O-RAN},'' in
  \emph{IEEE International Conference on Communications (ICC)}, 2023.

\bibitem{OranWG4}
{O-RAN Working Group 4}, ``O-{RAN} {F}ronthaul {C}ontrol, {U}ser and
  {S}ynchronization {P}lane {S}pecification v12,'' O-RAN.WG4.CUS.0-R003-v12.00,
  Tech. Rep., June 2023.

\bibitem{upadhyaya2022prototyping}
P.~S. Upadhyaya, A.~S. Abdalla, V.~Marojevic, J.~H. Reed, and V.~K. Shah,
  ``Prototyping {N}ext-{G}eneration {O-RAN} {R}esearch {T}estbeds with
  {SDR}s,'' \emph{arXiv preprint arXiv:2205.13178}, 2022.

\bibitem{OranWG11-secreqspec}
{O-RAN Working Group 11}, ``{Security Requirements Specifications},''
  O-RAN.WG11.Security-Requirements-Specification.O-R003-v06.00, Tech. Rep.,
  June 2023.

\bibitem{OranWG11}
------, ``{Security Protocols Specifications},''
  ORAN.WG11.Security-Protocols-Specifications-v04.00, Tech. Rep., July 2022.

\bibitem{OranWG4-M}
{O-RAN Working Group 4}, ``{O-RAN Management Plane Specification 12.0},''
  O-RAN.WG4.MP.0-R003-v12.00, Tech. Rep., June 2023.

\bibitem{OranWG5}
{O-RAN Working Group 5}, ``Transport specification,''
  ORAN.WG5.Transport.0-v01.00, Tech. Rep., March 2020.

\bibitem{tls1.3}
\BIBentryALTinterwordspacing
E.~Rescorla, ``{The Transport Layer Security (TLS) Protocol Version 1.3},'' Aug
  2018. [Online]. Available: \url{https://www.rfc-editor.org/rfc/rfc8446.txt}
\BIBentrySTDinterwordspacing

\bibitem{ipsec_ah}
\BIBentryALTinterwordspacing
``{IP Authentication Header},'' Dec 2005. [Online]. Available:
  \url{https://www.rfc-editor.org/rfc/rfc4303.txt}
\BIBentrySTDinterwordspacing

\bibitem{ipsec_esp}
\BIBentryALTinterwordspacing
``{IP Encapsulating Security Payload (ESP)},'' Dec 2005. [Online]. Available:
  \url{https://www.rfc-editor.org/rfc/rfc4303.txt}
\BIBentrySTDinterwordspacing

\bibitem{frankel2005guide}
S.~Frankel, K.~Kent, R.~Lewkowski, A.~D. Orebaugh, R.~W. Ritchey, and S.~R.
  Sharma, ``Guide to {IP}sec {VPN}s:.'' \emph{NIST Special Publication}, 2005.

\bibitem{macsec}
``{IEEE Standard for Local and metropolitan area networks-Media Access Control
  (MAC) Security},'' \emph{IEEE Std 802.1AE-2018 (Revision of IEEE Std
  802.1AE-2006)}, pp. 1--239, 2018.

\bibitem{802.1x}
``{IEEE Standard for Local and Metropolitan Area Networks--Port-Based Network
  Access Control},'' \emph{IEEE Std 802.1X-2020 (Revision of IEEE Std
  802.1X-2010 Incorporating IEEE Std 802.1Xbx-2014 and IEEE Std
  802.1Xck-2018)}, pp. 1--289, 2020.

\bibitem{kurose1986computer}
J.~F. Kurose and K.~W. Ross, \emph{Computer {N}etworking: {A} {T}op-{D}own
  {A}pproach, {S}eventh {E}ddition}, 2017.

\bibitem{project-strongswan}
\BIBentryALTinterwordspacing
strongSwan, ``Strongswan.'' [Online]. Available:
  \url{https://www.strongswan.org/}
\BIBentrySTDinterwordspacing

\bibitem{7784887}
V.~K. Choyi, A.~Abdel-Hamid, Y.~Shah, S.~Ferdi, and A.~Brusilovsky, ``Network
  slice selection, assignment and routing within 5{G} {N}etworks,'' in
  \emph{2016 IEEE Conference on Standards for Communications and Networking
  (CSCN)}, 2016, pp. 1--7.

\bibitem{groen2023tractor}
J.~Groen, M.~Belgiovine, U.~Demir, B.~Kim, and K.~Chowdhury, ``{TRACTOR:
  Traffic Analysis and Classification Tool for Open RAN},'' \emph{arXiv
  preprint arXiv:2312.07896}, 2023.

\bibitem{stewart2007stream}
R.~Stewart, ``Stream control transmission protocol ({RFC} 4960),'' Tech. Rep.,
  2007.

\bibitem{AES}
M.~Dworkin, E.~Barker, J.~Nechvatal, J.~Foti, L.~Bassham, E.~Roback, and
  J.~Dray, ``\BIBforeignlanguage{en}{Advanced {E}ncryption {S}tandard
  ({AES})},'' 2001-11-26 2001.

\bibitem{fips2012180}
Q.~Dang, ``\BIBforeignlanguage{en}{Secure hash standard},'' 2015-08-04 2015.

\bibitem{aerialsdk}
A.~Kelkar and C.~Dick, ``{NVIDIA Aerial GPU Hosted AI-on-5G},'' in \emph{IEEE
  4th 5G World Forum (5GWF)}, October 2021, pp. 64--69.

\bibitem{oai}
\BIBentryALTinterwordspacing
{OpenAirInterface Software Alliance}. [Online]. Available:
  \url{https://openairinterface.org}
\BIBentrySTDinterwordspacing

\bibitem{oneplus}
\BIBentryALTinterwordspacing
\emph{{OnePlus}}. [Online]. Available:
  \url{https://www.oneplus.com/us/nord-specs}
\BIBentrySTDinterwordspacing

\bibitem{Dubroca_2019}
\BIBentryALTinterwordspacing
S.~Dubroca, ``{manpage: IP-macsec - macsec device configuration},'' 2019.
  [Online]. Available: \url{https://manpages.ubuntu.com/
  manpages/jammy/en/man8/ip-macsec.8.html}
\BIBentrySTDinterwordspacing

\bibitem{mcgrew2004galois}
D.~McGrew and J.~Viega, ``{The Galois/counter mode of operation (GCM)},''
  \emph{submission to NIST Modes of Operation Process}, vol.~20, pp.
  0278--0070, 2004.

\bibitem{strongswanIKEv2Cipher}
\BIBentryALTinterwordspacing
``{I}{K}{E}v2 {C}ipher {S}uites :: strong{S}wan {D}ocumentation,'' 2023.
  [Online]. Available: \url{https://docs.strongswan.org/docs/5.9/config/
  IKEv2CipherSuites.html}
\BIBentrySTDinterwordspacing

\bibitem{ullah2020ipsec}
S.~Ullah, J.~Choi, and H.~Oh, ``{IPsec for high speed network links:
  Performance analysis and enhancements},'' \emph{Future Generation Computer
  Systems}, vol. 107, pp. 112--125, 2020.

\bibitem{7110118}
S.~Gallenmüller, P.~Emmerich, F.~Wohlfart, D.~Raumer, and G.~Carle,
  ``Comparison of frameworks for high-performance packet {IO},'' in \emph{2015
  ACM/IEEE Symposium on Architectures for Networking and Communications Systems
  (ANCS)}, 2015, pp. 29--38.

\end{thebibliography}

\begin{IEEEbiography}[{\includegraphics[width=1in,height=1.25in,clip,keepaspectratio]{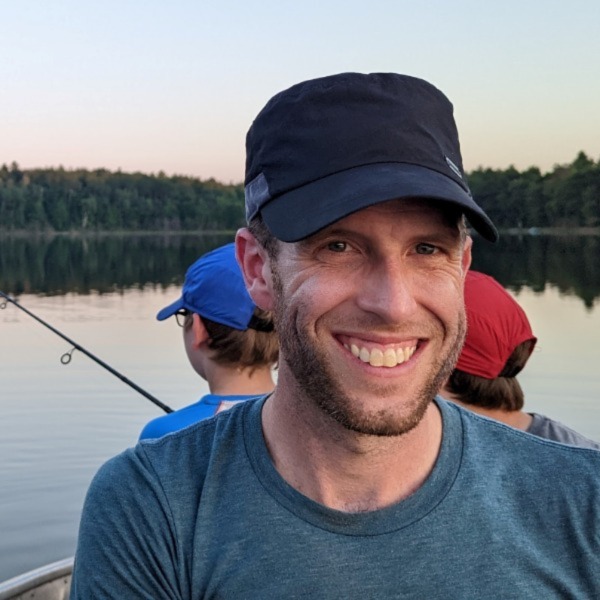}}]%
{Joshua Groen} is a Ph.D. candidate at Northeastern University. Previously he worked in the US Army Regional Cyber Center – Korea as the Senior Network Engineer. He received his BSE ('07) and MS ('17) in Electrical Engineering respectively from Arizona State University and the University of Wisconsin. His research interests include wireless communications, security, and machine learning.
\end{IEEEbiography}

\vspace{-0.5cm}
\begin{IEEEbiography}[{\includegraphics[width=1in,height=1.25in,clip,keepaspectratio]{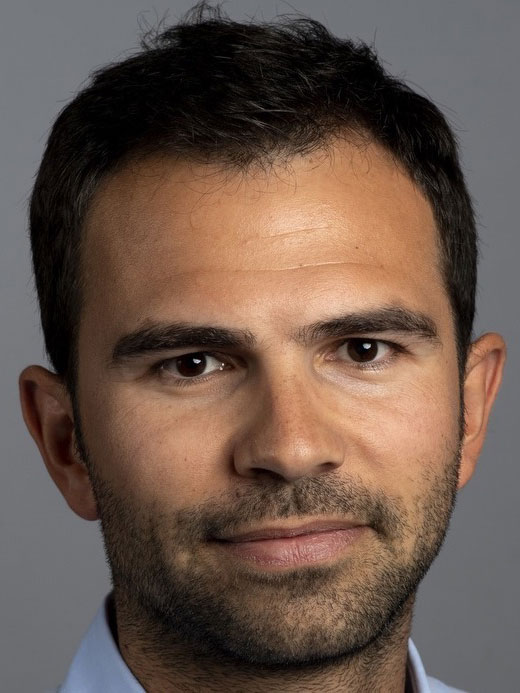}}]%
{Salvatore D’Oro} is a Research Assistant Professor at Northeastern University. He received his Ph.D. from the University of Catania in 2015. He serves on the Technical Program Committee of IEEE INFOCOM. His research focuses on optimization and learning for NextG systems.
\end{IEEEbiography}

\vspace{-0.5cm}
\begin{IEEEbiography}[{\includegraphics[width=1in,height=1.25in,clip,keepaspectratio]{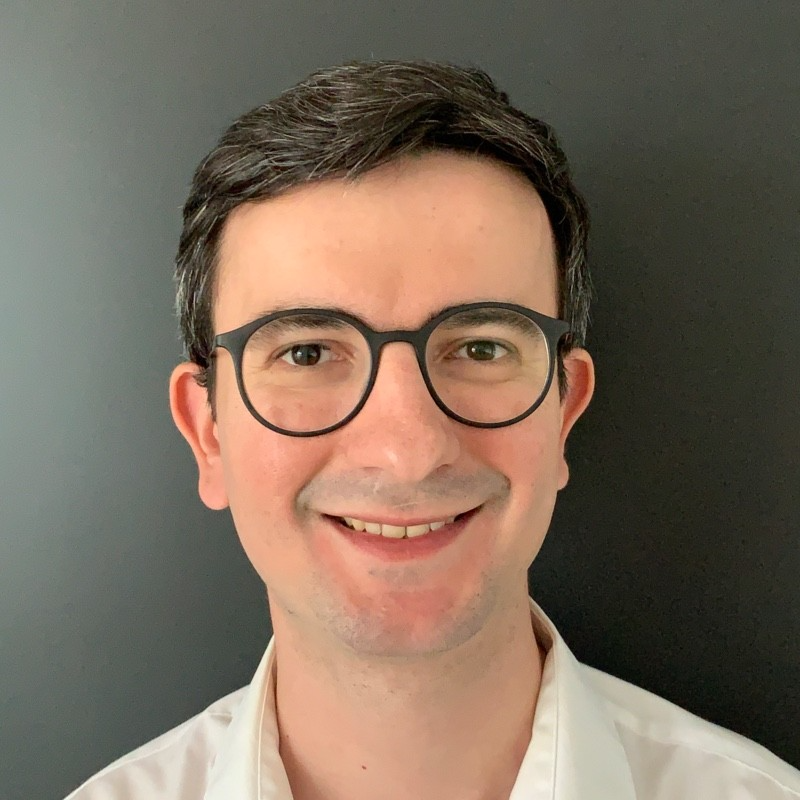}}]%
{Utku Demir} is a Postdoctoral Research Fellow at Northeastern University, where he is supported by the Roux Institute's Experiential AI program. He received his PhD from the University of Rochester in 2020. His research interests lie in the areas of wireless communications, mobile networks, signal processing, and machine learning.
\end{IEEEbiography}

\vspace{-0.5cm}
\begin{IEEEbiography}[{\includegraphics[width=1in,height=1.25in,clip,keepaspectratio]{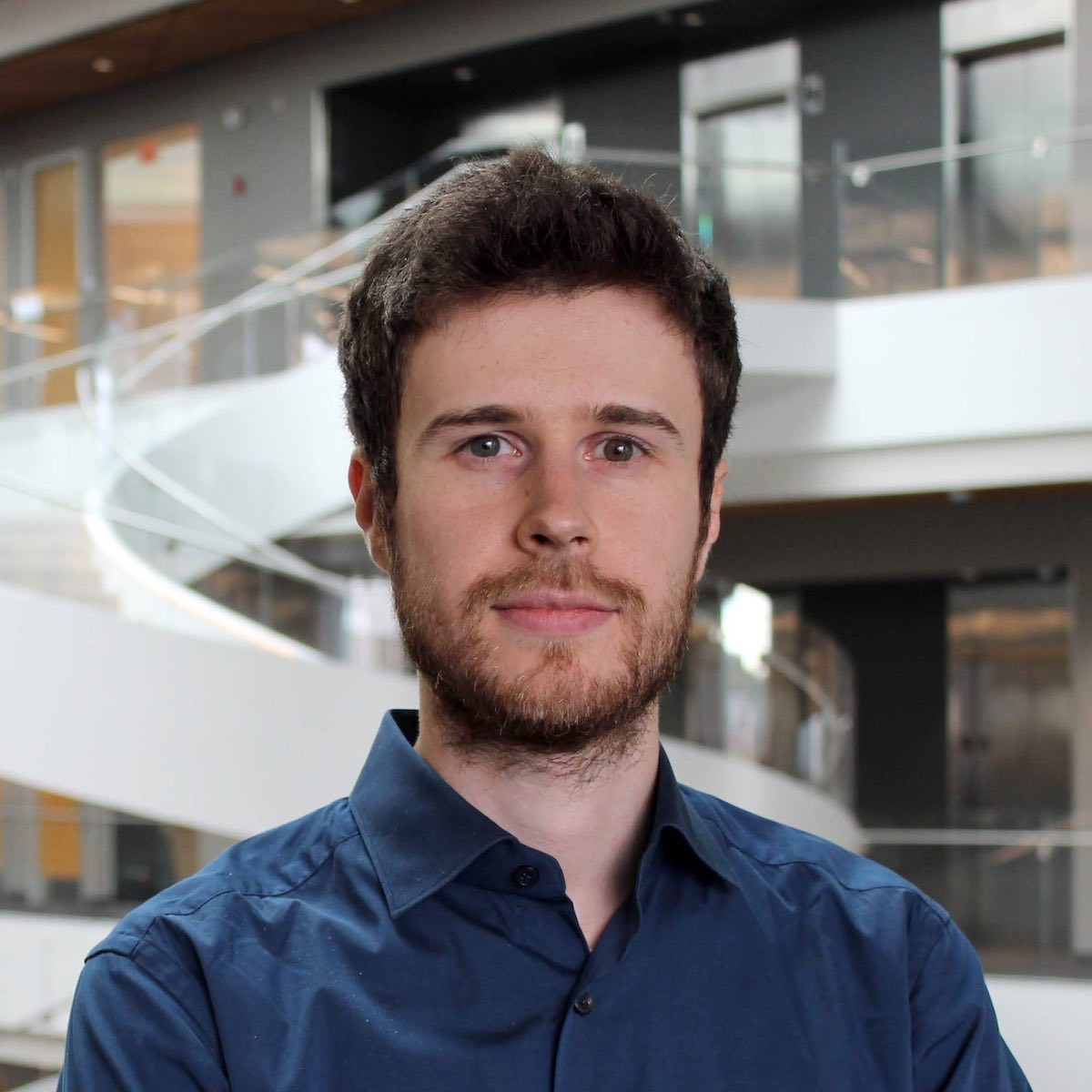}}]%
{Leonardo Bonati} is an Associate Research Scientist at Northeastern University. He received his Ph.D. from Northeastern University in 2022. His research focuses on softwarized NextG systems.
\end{IEEEbiography}

\vspace{-0.5cm}
\begin{IEEEbiography}[{\includegraphics[width=1in,height=1.25in,clip,keepaspectratio]{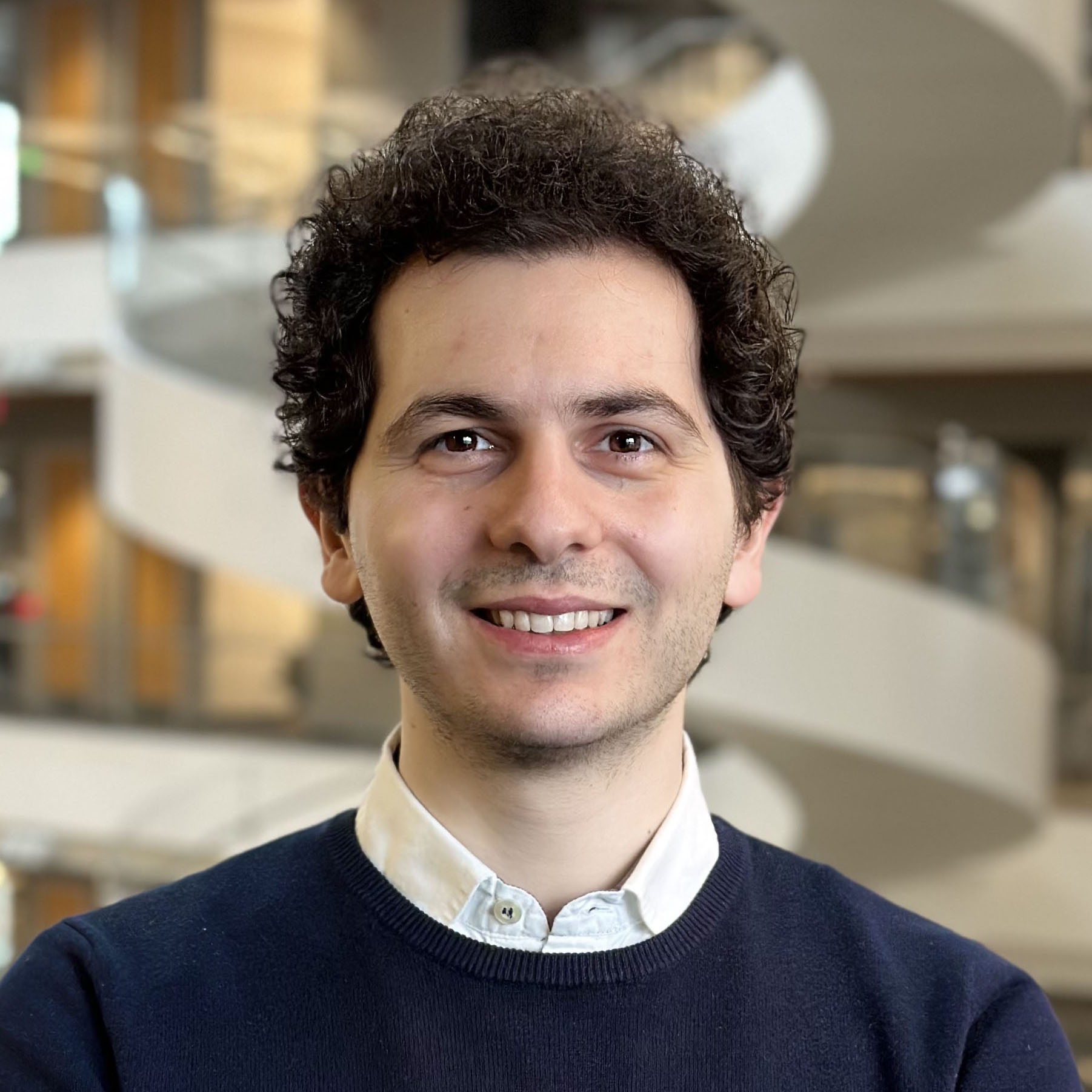}}]%
{Davide Villa} is a Ph.D. candidate at Northeastern University. He received his B.S. in Computer Engineering and his M.S. in Embedded Computing Systems from University of Pisa and Sant'Anna School of Advanced Studies in 2015 and 2018, respectively. His research interests focus on 5G-and-beyond cellular networks, software-defined networking, O-RAN, and channel modeling.
\end{IEEEbiography}

\vspace{-0.5cm}
\begin{IEEEbiography}[{\includegraphics[width=1in,height=1.25in,clip,keepaspectratio]{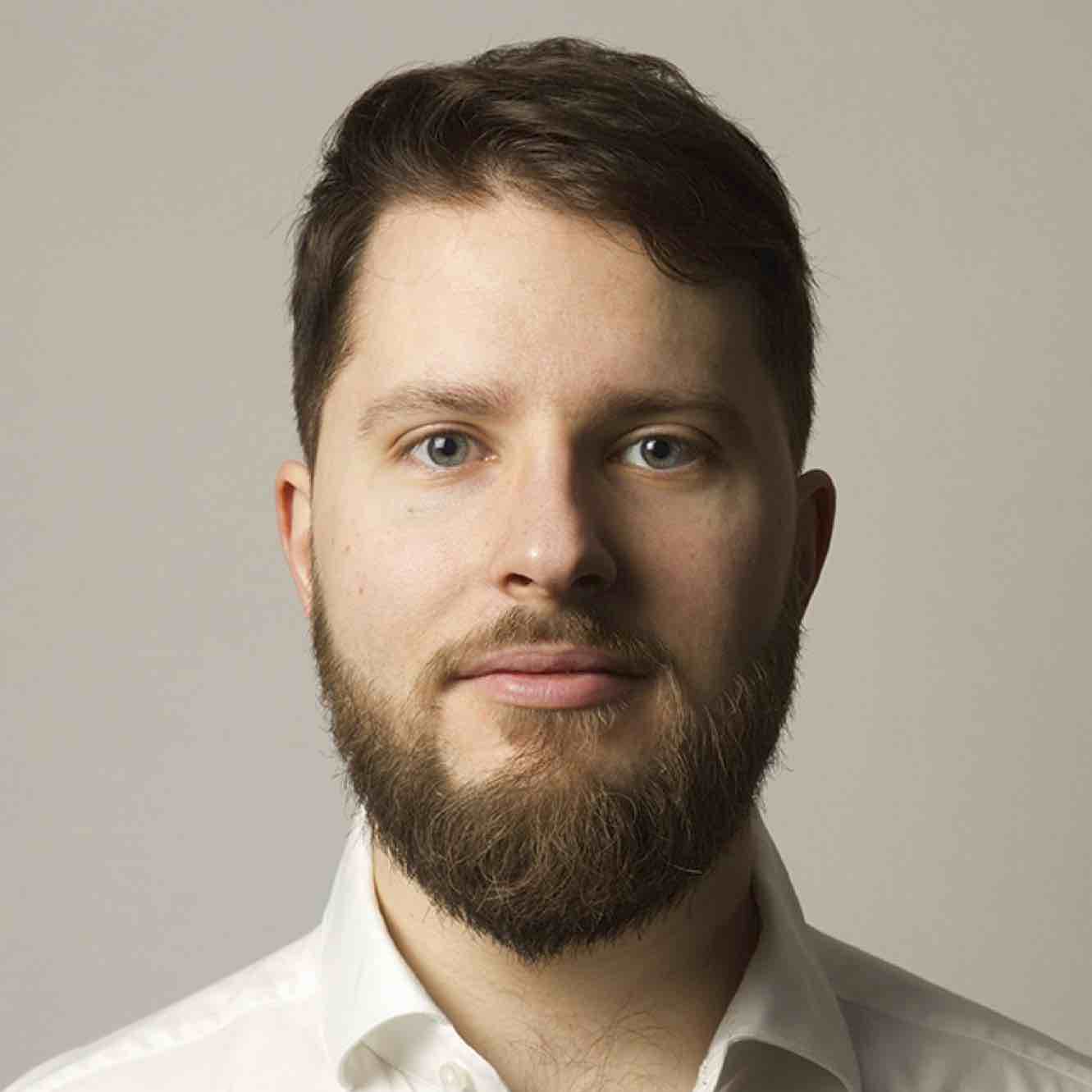}}]%
{Michele Polese} is a Research Assistant Professor at Northeastern University. He obtained his Ph.D. from the University of Padova in 2020. His research focuses on architectures for wireless networks.
\end{IEEEbiography}

\vspace{-0.5cm}
\begin{IEEEbiography}[{\includegraphics[width=1in,height=1.25in,clip,keepaspectratio]{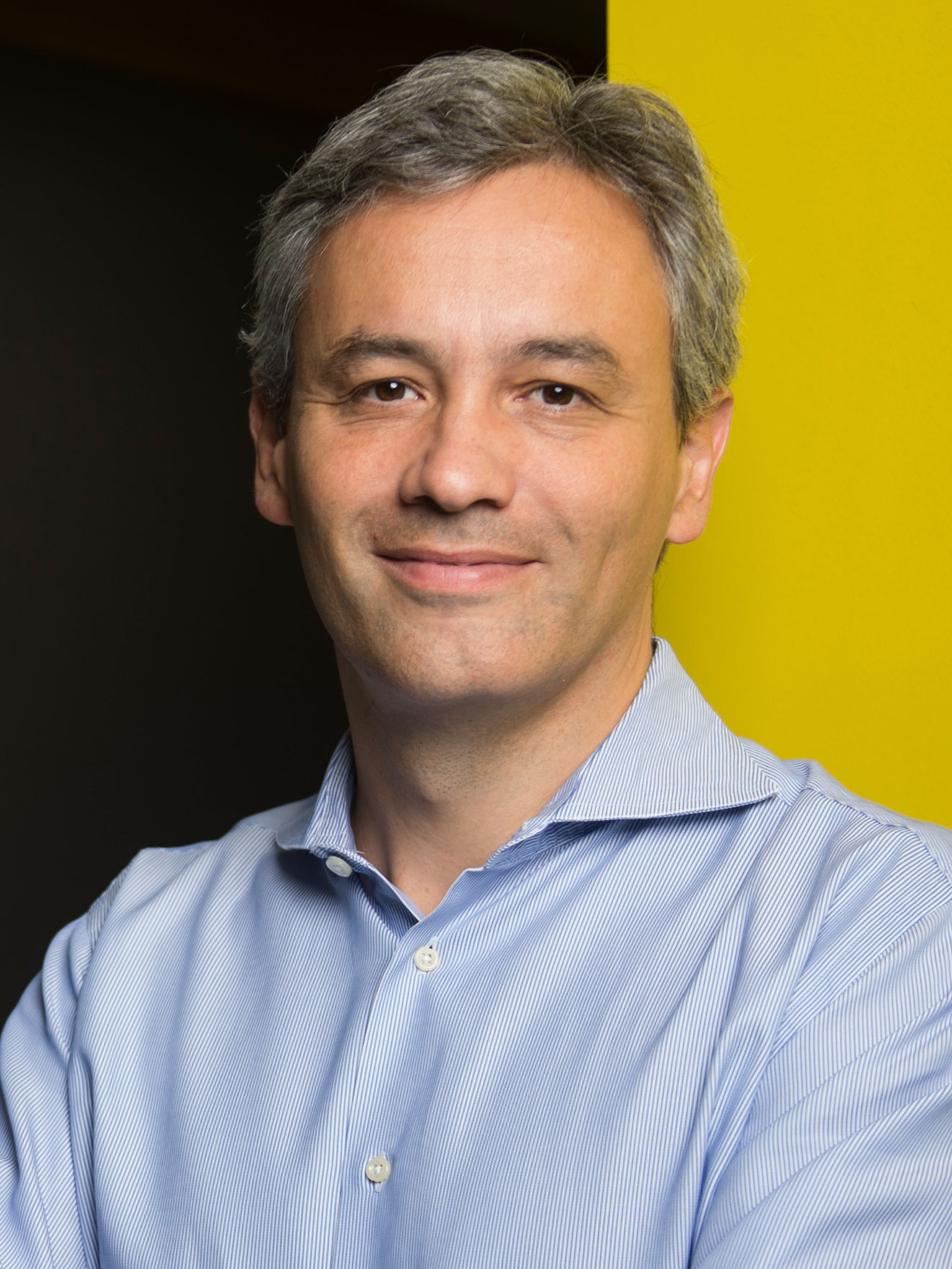}}]%
{Tommaso Melodia} is the William Lincoln Smith Professor at Northeastern University, the director of the Institute for the Wireless Internet of Things, and the director of research for the PAWR Project Office. He received his PhD from the Georgia Institute of Technology in 2007. His research focuses on wireless networked systems.
\end{IEEEbiography}

\vspace{-0.5cm}
\begin{IEEEbiography}[{\includegraphics[width=1in,height=1.25in,clip,keepaspectratio]{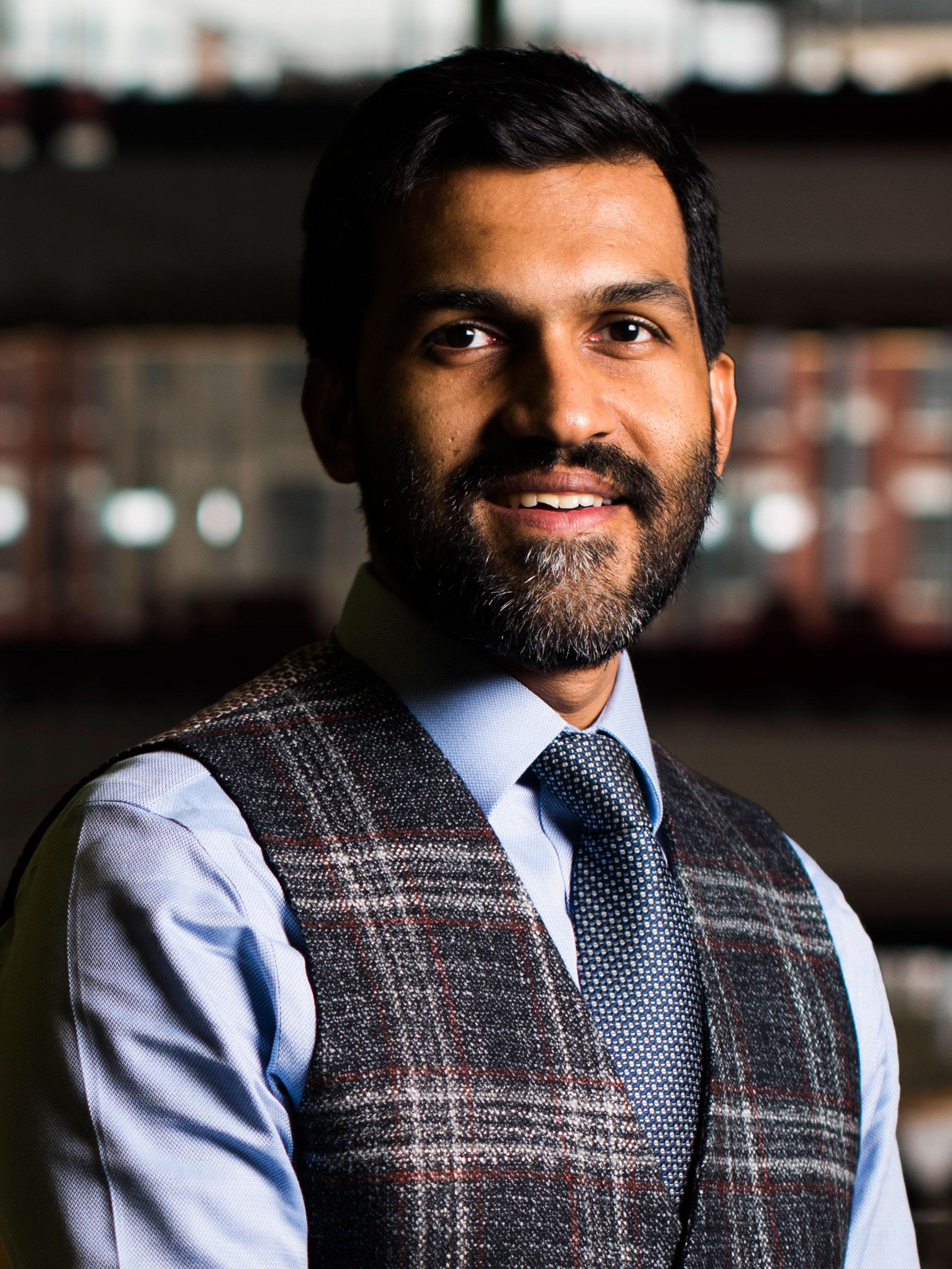}}]%
{Kaushik Chowdhury} is a Professor at Northeastern University, Boston, MA. He received his PhD from Georgia Institute of Technology in 2009. His current research interests involve systems aspects of machine learning for agile spectrum sensing/access, unmanned autonomous systems, programmable and open cellular networks, and large-scale experimental deployment of emerging wireless technologies.
\end{IEEEbiography}

\end{document}